%% file: ebintro_paper.tex
\documentclass{aastex62}

\usepackage{graphicx}
\usepackage{enumerate}

\usepackage{url}
\usepackage{color}
\usepackage{upgreek}
\usepackage{longtable}
\usepackage{mathrsfs}
\usepackage{kellymacros2}
\usepackage{natbib}

\newcommand{\um}{\ensuremath{\mathrm{\,\mu{}m}}}
\newcommand{\msun}{\ensuremath{\mathit{\,M_\odot}}}
\newcommand{\rsun}{\ensuremath{\mathit{\,R_\odot}}}
\newcommand{\kps}{\ensuremath{\mathrm{\,km\,s^{-1}}}}
\newcommand{\mps}{\ensuremath{\mathrm{\,m\,s^{-1}}}}
\newcommand{\Kepler}{\emph{Kepler}}

\newcommand{\e}{\ensuremath{\mathit{e}}}

\newcommand{\w}{\ensuremath{\mathit{\omega}}}

\newcommand{\q}{\ensuremath{\mathit{q}}}
\newcommand{\KA}{\ensuremath{\mathit{K}_A}}
\newcommand{\KB}{\ensuremath{\mathit{K}_B}}

\newcommand{\g}{\ensuremath{\mathit{\gamma}}}

\newcommand{\MA}{\ensuremath{\mathit{M}_A}}
\newcommand{\MB}{\ensuremath{\mathit{M}_B}}
\newcommand{\RA}{\ensuremath{\mathit{R}_A}}
\newcommand{\RB}{\ensuremath{\mathit{R}_B}}

\newcommand{\TA}{\ensuremath{\mathit{T}_A}}
\newcommand{\TB}{\ensuremath{\mathit{T}_B}}

\shorttitle{SDSS HET \Kepler EBs}
\shortauthors{Mahadevan et al.}

\begin{document}


\title{The SDSS-HET Survey of \Kepler{} Eclipsing Binaries. Description of the Survey and First Results.}


\author{Suvrath Mahadevan}
\affil{Department of Astronomy \& Astrophysics,  The Pennsylvania State University, 525 Davey Lab, University Park, PA-16802}
\affil{Center for Exoplanets $\&$ Habitable Worlds, The Pennsylvania State University, 525 Davey Lab, University Park, PA-16802}
\email{suvrath@astro.psu.edu}

\author{Chad F. Bender}
\affil{Department of Astronomy and Steward Observatory, University of Arizona, Tucson, AZ 85721}

\author{Kelly Hambleton}
\affil{Department of Astronomy \& Astrophysics, Villanova University, 800 East Lancaster Ave., Villanova, PA 18085}

\author[0000-0003-0556-027X]{Scott W. Fleming}
\affil{Space Telescope Science Institute, 3700 San Martin Dr, Baltimore, MD 21218}

\author{Rohit Deshpande}
\affil{Department of Astronomy \& Astrophysics,  The Pennsylvania State University, 525 Davey Lab, University Park, PA-16802}
\affil{Center for Exoplanets $\&$ Habitable Worlds, The Pennsylvania State University, 525 Davey Lab, University Park, PA-16802}

\author{Kyle Conroy}
\affil{Department of Astronomy \& Astrophysics, Villanova University, 800 East Lancaster Ave., Villanova, PA 18085}

\author{Gal Matijevi\v c}
\affil{Department of Astronomy \& Astrophysics, Villanova University, 800 East Lancaster Ave., Villanova, PA 18085}

\author{Leslie Hebb}
\affil{Physics Department, Hobart and William Smith Colleges, 300 Pulteney Street, Geneva, NY 14456}

\author{Arpita Roy}
\affil{California Institute of Technology, 1200 E California Blvd, Pasadena, CA 91125}

\author{Hasan Ak}
\affil{Erciyes University, Science Faculty, Astronomy and Space Sci. Dept., 38039 Kayseri, Turkey}

\author{Bla\v z Leban}
\affil{University of Ljubljana, Dept.~of Physics, Jadranska 19, SI-1000 Ljubljana, Slovenia}
\affil{Department of Astronomy \& Astrophysics, Villanova University, 800 East Lancaster Ave., Villanova, PA 18085}

\author[0000-0002-1913-0281]{Andrej Pr\v{s}a}
\affil{Department of Astronomy \& Astrophysics, Villanova University, 800 East Lancaster Ave., Villanova, PA 18085}
\email{aprsa@villanova.edu}

\begin{abstract}
The \Kepler{} mission has provided a treasure trove of eclipsing binaries (EBs), observed at extremely high photometric precision, nearly continuously for several years. We are carrying out a survey of $\sim$100 of these EBs to derive dynamical masses and radii with precisions of 3\% or better.  We use multiplexed near-infrared H band spectroscopy from the Sloan Digital Sky Survey-III and -IV APOGEE instrument and optical spectroscopy from the Hobby Eberly telescope High-Resolution Spectrograph to derive double-lined spectroscopic orbits and dynamical mass-ratios (\q) for the EB sample, two of which we showcase in this paper.  This orbital information is combined with \Kepler{} photometry to derive orbital inclination, dynamical masses of the system component, radii and temperatures.  These measurements are directly applicable for benchmarking stellar models that are integrating the next generation of improvements, such as the magnetic suppression of convection efficiency, updated opacity tables, and fine-tuned equations of state.  We selected our EB sample to include systems with low-mass ($M\la0.8\,\msun$) primary or secondary components, as well as many EBs expected to populate the relatively sparse parameter space below $\sim0.5\,M_{\odot}$.  In this paper, we describe our EB sample and the analysis techniques we are utilizing, and also present masses and radii for two systems that inhabit particularly underpopulated regions of mass-radius-period space: KIC 2445134 and KIC 3003991. Our joint spectrosopic and photometric analysis of KIC 2445134 ($\q{}=0.411\pm0.001$) yields masses and radii of $\MA{}=1.29\pm0.03\msun$, $\MB{}=0.53\pm0.01\msun$, $\RA{}=1.42\pm0.01\rsun$, $\RB{}=0.510\pm0.004\rsun$, and a temperature ratio of $T_B/T_A=0.635\pm0.001$; our analysis of KIC 3003991 ($\q{}=0.298\pm0.006$) yields $\MA{}=0.74\pm0.04\msun$, $\MB{}=0.222\pm0.007\msun$, $\RA{}=0.84\pm0.01\rsun$, $\RB{}=0.250\pm0.004\rsun$, and a temperature ratio of $T_B/T_A=0.662\pm0.001$.

\end{abstract}



\section{Introduction}
Eclipsing binaries (EBs) have served as benchmarks for stellar
astrophysics for hundreds of years
\citep[e.g.,][]{1783RSPT...73..474G}.  Advances in radial velocity
(RV) measurement precision over the past several decades have steadily
improved the precision of masses derived from EBs. Compilations of EBs
show that the total number, diversity, and measurement precisions of
detached, main-sequence EBs have steadily increased: 72 stars with
masses and radii ($M$, $R$) measured to $<$ 15\% in
\citet{1980ARA&A..18..115P}, 88 stars with $M$, $R$ $<$ 2\% in
\citet{1991A&ARv...3...91A}, 188 stars with $M$, $R$ $<$ 3\% in
\citet{2010A&ARv..18...67T}, and 198 stars with $M$, $R$ $<$2\% in
DEBCat\footnote{\url{http://www.astro.keele.ac.uk/jkt/debcat/}}, an
updated version of \citet{1991A&ARv...3...91A}.

Despite this steady growth in sample size, low-mass stars (defined
here as $M < 0.8\,M_{\odot}$) remain a relatively small fraction of
the overall sample. For example, DEBCat contains only 54 stars with $M
< 0.8\,M_{\odot}$, while the \citet{2010A&ARv..18...67T} sample has
only 10 such stars, none of which have orbital periods longer than five
days. More recent results are starting to expand the low-mass sample
\citep[e.g.,][]{Schwamb2013,Gomez2014,Zhou2015,Dittmann2017,Lubin2017, Casewell2018}, but a significant increase
in sample size is still lacking, predominantly due to the resource
expense associated with obtaining high-precision RVs (the method used
to derive most of the EB dynamical masses in the current
compilations), and because the poor flux ratio of a low-mass K/M dwarf
orbiting a larger primary makes it difficult to detect in the optical.

Observations in the near-infrared (NIR) result in the tangible benefit
of improving the flux contrast for EBs with a small mass-ratio
($\q=\MB/\MA$). For example, the flux ratio of an M5 dwarf, with
$T_{\rm{eff}}\sim3300$K to a G2 dwarf, with $T_{\rm{eff}}\sim5800$K, is $\sim10$
times more favorable in the $H$ band compared to the $V$ band.

This allows for the extraction of measured RVs
from fainter secondary stars, and therefore extends the lower limit on
\q{} for which masses of dwarf stars can be derived. While optical
spectra tend to lose sensitivity at $q \sim 0.5$, NIR spectra can
push down to $q \sim 0.1$
\citep{Prato2002,2003ApJ...599.1344M,2008ApJ...689..416B}.

Measurements of select bright binaries have achieved mass ratio
precisions of 0.02-0.42\% (in $M\sin{i}$) \citep{2010ApJ...719.1293K},
but precisions of 1-3\% are still sufficient to distinguish between
model parameters \citep{2010ApJ...718..502M}. The high photometric
precision of \Kepler{} makes it possible to determine the masses of some
EBs purely photometrically \citep{2011Sci...331..562C,
  2012ApJ...746..185F} via Doppler boosting and ellipsoidal
effects \citep{2011MNRAS.415.3921F}. While the technique is promising,
\citet{2011Sci...331..562C} achieved only a 10\% mass precision using
the \Kepler{} photometry alone (i.e., without including any
spectroscopic RVs), while \citet{2012ApJ...746..185F} found that only
5 of their 7 binaries had RV semi-amplitudes in agreement with those
predicted from their photometric analysis.  In a large follow-up program, \citet{talor2015} observed 281 targets, confirming 70 binary systems while finding many of their false positives were due to pulsating red giants.  The use of spectroscopic
RVs therefore remains an essential and reliable method for
obtaining mass precisions at the 1-3\% level, motivating the continual
use of spectra to measure the stellar masses of EBs.

Stellar models rely on the precision measurements that EBs afford for
calibrating the physical parameters used in their calculations
(leading to well-characterized EBs often being referred to as stellar
``benchmarks''). Stars with $M > 0.8\,M_{\odot}$ generally agree with
theoretical models to within observational uncertainties. Lower mass
stars, however, are often observed to have radii that are 5-15\%
larger than model predictions
\citep{2007ApJ...660..732L,2008A&A...478..507M,2010A&ARv..18...67T,2011ApJ...728...48K,Higl2017,Cruz2018,Kesseli18}. There
is observational and theoretical evidence to suggest that magnetic
fields could be the cause
\citep{2007A&A...472L..17C,2010ApJ...718..502M,2013ApJ...779..183F,MacDonald2017},
due to interaction with the partially convective outer atmospheres
and/or generation of cool starspots at polar latitudes. Unidentified
opacity sources have also been suggested \citep{2006ApJ...644..475B},
although a lack of metallicity measurements for many stars in the
current EB sample has prevented an in-depth examination of that
possible correlation. Models that use a different equation
of state than previous works agree with observations of one low-mass
\Kepler{} EB \citep[KOI-126;][]{2011ApJ...740L..25F}, opening another
regime of parameter space to explore.

In this paper we introduce the \emph{SDSS-HET Survey of Kepler
  Eclipsing Binaries}, which is combining \Kepler{} photometry with
ground-based spectroscopy to precisely measure orbital parameters,
dynamical masses and radii, and temperature ratios for a sample of 109
EB candidates selected from the \citet{Kirk2016} catalog, and listed in Table~\ref{ebsampletable}.
We utilize both optical and NIR spectroscopy to solve these EBs as
double-lined spectroscopic binaries (SB2s), and combine these results
with \Kepler{} photometry to derive masses and radii with precisions
of better than 3\% for most of the sample, and as good as 1\% for a
subset. Here we present two objects in our sample.  Our complete sample is restricted to EBs that are classified as fully
detached and have $H<13$.  Orbital periods range from a few days to
more than one hundred days.  The total sample size is comparable in
number to the $M<2\msun$ members in \citet{2010A&ARv..18...67T} and
DEBCat (110 and 285 EBs, respectively),
but we have included many EBs with low-mass primary or secondary
components in order to substantially increase the population of well
measured low-mass stars.  One of the reasons the compilation from
\citet{2010A&ARv..18...67T} is so useful to the astronomical community
is that their study homogenously re-computed orbital and stellar
parameters from the compiled list of literature EBs.  Our survey will
intrinsically possess this quality, as our analysis applies a singular
set of tools to a homogeneous data set.

 In \S2 we describe the facilities and data
products we are using, and in \S3 we discuss the analysis techniques applied
to the spectroscopy and photometry, as well as probabilistic analysis that we use to derive realistic parameter uncertainties.  In
\S4 we present two low-mass EBs from our sample, and provide our
derived orbits, masses, and radii for these systems.  In \S5 we
discuss these systems in the context of the pre-existing population of
precisely measured low-mass EBs, and describe our plans for analyzing
the remainder of our \Kepler{} EB sample.  As part of this program, we
constructed a semi-automated reduction pipeline for the HET
High-Resolution Spectrograph, which we describe in Appendix~A.

\section{Facilities and Datasets\label{facilities}} 

\subsection{The \Kepler{} Mission\label{kepler}}

\Kepler{} is a space-borne, 0.95-m, high-precision photometer equipped
with a broadband filter covering 420 nm -- 865 nm.  From 2009 -- 2013,
\Kepler{} monitored a single field located $13.5^{\circ}$ above the
galactic plane in the direction of Cygnus, with a mission to detect
transiting, habitable-zone exoplanets \citep{2010Sci...327..977B}. Its
ability to conduct photometry with high precision \citep[$\sim80$ ppm
  for $K_p = 12$;][]{2010ApJ...713L..92C} and to observe with a nearly
continuous cadence facilitated numerous ancillary stellar astrophysics
programs, including the detection of EBs over a range of orbital
periods that are challenging to observe from the ground. The majority of \Kepler{} EBs are faint \citep[only $\sim12$\%
  of detached systems have $K_p < 12$ in][]{Kirk2016}, and
followup spectroscopy of such systems requires a combination of large
telescopes and long exposure times. The
\emph{Kepler Eclipsing Binary
  Catalog}\footnote{http://keplerebs.villanova.edu/} \citep{2005ApJ...628..426P,2011AJ....142..160S,Matijevic2012,Conroy2014,Kirk2016,Abdul2016} has used
\Kepler{} photometry to identify thousands of EBs, and has derived
extremely precise orbital periods, which greatly simplifies the
process of turning individual spectroscopic measurements into an SB2
orbit. 

\subsection{The SDSS-III APOGEE Spectrometer\label{apogee}}

The Apache Point Observatory Galactic Evolution Experiment (APOGEE, \citealp{Majewski2017})
is a fiber-fed, multi-object, near-infrared spectrometer that uses a
volume phase holographic grating and a linear array of three
Hawaii-2RG detectors to record spectra from 1.51\um{} to 1.68\um{}
with a spectral resolution of $\lambda/\Delta\lambda\sim22,500$
\citep{2010SPIE.7735E..46W}. The instrument is located at Apache Point
Observatory on the 2.5m SDSS telescope \citep{2006AJ....131.2332G,York2000},
and was commissioned in the spring and summer of 2011 for a three year
survey aimed at a galactic evolution experiment \citep{2008AN....329.1018A,2011AJ....142...72E}.  A
small fraction of the survey time ($\sim5\%$) was devoted to ancillary
science programs \citep{2013AJ....146...81Z}, which includes the
\Kepler{} EB program described in this paper.  The spectrograph is
stabilized in a vacuum-sealed cryostat cooled via liquid nitrogen,
which minimizes thermal variations and yields a typical radial
velocity precision of $100-200\mps$ on our EB sample.  APOGEE can
simultaneously observe 300 targets over the telescope's 3$^\circ$
diameter field-of-view, which is also fortuitously the approximate size of each
\Kepler{} module. This multiplexing capability allows us to
efficiently observe many EBs with a single integration, also referred
to as a \emph{visit}.  Each \emph{visit} is typically comprised of eight
consecutive eight-minute exposures, which are later combined by the
APOGEE data reduction pipeline \citep{Nidever2015} to produce a
single \emph{visit} spectrum with a total integration time of slightly
more than one hour.  All analysis described in this paper utilizes
APOGEE data at the \emph{visit} level of processing.

Our ancillary APOGEE program included two fields overlapping \Kepler{}
modules centered on the open clusters NGC 6791 (3 \emph{visits}) and
NGC 6819 (6 \emph{visits}).  We observed a total of 42 and 67 detached
EBs in each field, respectively.  All spectra for this program were
obtained in 2011, during the first year of survey operations, and have been
publicly released as part of SDSS Data Release 10
\citep{2013arXiv1307.7735A}. 

Prior to analyzing the APOGEE spectra, we perform additional
post-processing beyond that provided by the APOGEE pipeline.  We use a
low order polynomial to remove continuum and normalize each spectrum.
Residuals caused by imperfect correction of telluric absorption and
sky emission are present in most pipeline reduced APOGEE spectra; we
manually correct these by interpolating over neighboring pixels.  The
pipeline version used for the DR10 release flagged wavelengths with
suspected bad pixels by setting their flux to zero.  This flagging
complicates our RV measurements because a cross-correlation analysis
interprets such pixels as having strong, discrete signal that is not
present in correlation templates, which reduces the overall amplitude
of a real correlation signal.  We interpolate over these regions to
minimize their impact.

\subsection{The HET High-Resolution Spectrograph\label{hrs}}

 The small number of visits
obtained by APOGEE in each of our fields are, by themselves, inadequate for deriving
stellar mass at the 3\% level for most of our EBs, but the infrared
bandwidth provides essential leverage on the low-mass companions in
high-contrast systems.  Consequently, for 55 of our EBs
(Table~\ref{ebsampletable}, column 7) we supplemented the APOGEE
spectroscopy with optical spectroscopy from the Hobby-Eberly
Telescope to achive higher precision in spectroscopic orbital determination for the primary star. The High-Resolution Spectrograph \citep[][hereafter
  HRS]{1998SPIE.3355..387T} is a visible light fiber-fed,
cross-dispersed, echelle spectrometer located on the 9.2\,m
Hobby-Eberly Telescope \citep[][hereafter HET]{1998SPIE.3352...34R}.
The HET was designed to carry out narrow-field spectroscopy of faint
objects, and so is well suited for targets like our \Kepler{} EB
sample.  The spectrograph is fiber-fed and housed in an isolated
enclosure in the HET spectrograph room, which provides a moderately
stable environment capable of achieving long-term radial velocity
stability of $25\mps$ on bright, low-mass, main-sequence stars using
standard ThAr emission lamps for wavelength calibration
\citep{2012ApJ...751L..31B}.  The HET operates under a queue-based
observing scheme \citep{2007PASP..119..556S} that allows us to request
observations of a target to occur within a narrow time window.  This
capability allowed us to efficiently observe 55 members of our EB
sample from 2011 -- 2013, while targeting each system at specific
orbital phases.  A similar set of observations could not easily be
obtained at a classically scheduled facility, making the HET a unique
resource for studying binary stars.

We use the HRS with a 2'' fiber in a configuration that provides a
spectral resolution of $\lambda / \Delta\lambda=30,000$ over a
bandwidth from 4076\,\AA -- 7838\,\AA, except a small gap at
5936\,\AA{} where light falls between the two HRS CCDs, distributed in 73
spectral orders.  This configuration provides a large number of
these features from which we derive precise radial velocities.  The
HET queue-based observing mode allows for the spectrograph
configuration to change multiple times throughout a single night 
as targeting requests from different
observers are carried out.  Standard sequences of calibration frames,
including \emph{biases}, \emph{flats}, and \emph{ThAr} wavelength
references, are typically obtained at the end of the observing
session.  The echelle and cross-disperser positions are not precisely
repeatable at the sub-pixel level, so the standard HET operations
introduce a discontinuity between the \emph{target} and the
\emph{ThAr} observations, which can result in RV shifts of several
hundred m s$^{-1}$.  To avoid this, we obtain additional \emph{ThAr}
frames immediately before or after each target observation, without
altering the instrument configuration.  We do not obtain extra
\emph{flats} because the telescope overhead would be severe and our experience has shown that the
misalignment inherent in the standard queue procedures is usually
small enough to prevent fringing in reduced images.

To efficiently and uniformly reduce the large HRS data set generated
by our \Kepler{} EB program, we created a semi-automated data handling
pipeline for the HRS.  This pipeline is written in the Interactive
Data language (IDL) and carries out image processing, spectral
extraction, and wavelength calibration tasks. Appendix A gives a detailed
description of the pipeline.  

As with the APOGEE spectra, we apply several post-processing steps to
our extracted HRS spectra prior to analysis.  Each spectral order is
continuum normalized.  Strong telluric contamination is mostly
restricted to isolated wavelength regions at optical wavelengths.
Rather than attempt a telluric correction across our HRS bandpass, we
choose to retain only those regions of the spectrum with telluric
contamination of 0.5\% or smaller, modulus a continuity function that
preserves large unbroken chunks of spectrum.  In practice, we retain
eight spectral windows: 4390--5025\,\AA, 5100--5410\,\AA,
5475--5680\,\AA, 5770--5855\,\AA, 6020--6260\,\AA, 6365--6430\,\AA,
6620--6850\,\AA, and 7450--7580\,\AA.  Finally, sky emission lines are
removed by interpolating over neighboring pixels.

\section{Analysis Techniques\label{analysis}}

\subsection{Measurement of Radial Velocities \label{rvanalysis}}

All spectroscopy of unresolved binary stars contain the blended light
of both the primary and secondary components.  The S/N and the
wavelength-dependent contrast ratio of an individual spectrum dictate
whether that spectrum can be solved as a double-lined spectroscopic
binary (SB2) or only as a single-lined spectroscopic binary (SB1).
Our APOGEE spectroscopy (\S\ref{apogee}) was designed to exploit the
NIR contrast advantage of EBs containing a pair of main-sequence stars
with a small \q, thereby observing the systems as SB2s.  Because our target sample contains a wide range of binary
types, including EBs with small \q{}, EBs with equal mass components,
and EBs with one or both components evolving or evolved, the
suitability of optical or NIR spectroscopy for solving each EB as an
SB2 is determined on a case by case basis.

We analyze our processed APOGEE and HET spectra identically, using a
combination of one-dimensional and two-dimensional cross-correlation
algorithms to measure RVs for the primary and secondary components of
each EB.  For the two-dimensional case, we have implemented the TODCOR
algorithm \citep{1994ApJ...420..806Z} as an interactive IDL program,
SXCORR, which allows the user great flexibility in quickly optimizing
the correlation templates while examining multiple epochs of spectra.
SXCORR simultaneously cross-correlates two template spectra against a
target spectrum containing the blended light from a binary to
disentangle the component RVs.  Our SXCORR implementation of TODCOR
includes the maximum-likelihood extension described by
\citet{2003MNRAS.342.1291Z}, modified slightly to allow segments of
spectrum with variable lengths.  SXCORR automatically resamples both
target and template spectra into log-lambda wavelength space
\citep{1979AJ.....84.1511T}, as needed depending on the input spectra.
Previous investigations have extensively described our procedures for
using TODCOR techniques to measure the RVs of an SB2
\citep[e.g.][]{2014arXiv1402.0846L, 2012ApJ...751L..31B,
  2008ApJ...689..416B}, and we refer the interested reader to the
discriptions therein.  Our one-dimensional correlation analysis is the
trivial simplification of the two-dimensional case.

Concurrently with our HRS EB observations, we observed an extensive
spectral template library of known single dwarf stars using an
identical HRS configuration.  This library ranges in spectral type
from early F through mid M.  Additionally, we have supplemented this
library with synthetic templates generated from the \texttt{PHOENIX}-based BT-Settl model grid \citep{2011ASPC..448...91A}.  These
synthetic models cover a much wider range of $T_{eff}$,
$\mathrm{[M/H]}$, and $\log{g}$ than our observed library, although some
demonstrate substantial discrepancies with observed spectra at
high-resolution \citep{2014ApJ...782...61T} that manifest in our RV
analysis as reduced correlation power.  To generate a template from
the BT-Settl library, we convolve the raw synthetic spectrum to the
proper resolution (22,500 for APOGEE; 30,000 for HRS), and resample to
3-pixels per resolution element.  We additionally apply a suite of
rotational broadening kernels generated from a four parameter
non-linear limb-darkening model \citep{2012A&A...546A..14C,
  1992oasp.book.....G} and the appropriate stellar parameters.  We do
not have an observed template library for the H-band APOGEE spectra,
so all APOGEE RVs are measured using BT-Settl templates.

\subsection{Binary Star Modeling}

To model the \Kepler{} light curve with {\sc {apogee}} and {\sc{het}} radial velocity data simultaneously, we used \ph\ 1.0 binary star modeling software \citep{2005ApJ...628..426P}, which is based on the Wilson-Devinney code \citep[][hereafter WD]{1971ApJ...166..605W}. \added{To make certain that there are no numerical or systematic artifacts arising from the choice of the legacy model, we synthesized light and RV curves using PHOEBE 2.1 \citep{prsa2016} and found no evidence for any discrepancy that exceeds the order of data scatter.} We fit the data within a Bayesian framework, by utilizing \emcee, a \python\ implementation of the affine invariant Markov chain Monte Carlo (MCMC) ensemble sampler proposed by \citet{Goodman2010} and implemented by \citet{2013PASP..125..306F}. We model the noise, instrumental variations and stellar variations caused by spots by employing \celerite, a Gaussian process library \citep{DFM2017}. Prior to modeling, we minimally prepared the data by flux-normalizing and stitching the \Kepler{} quarters and removing any obvious spurious points. The entire process is streamlined into the pipeline that is capable of processing the data autonomously. We describe the details below.

\subsubsection{Uncertainty Determination for the \kep\ Data}

As the \kep\ data uncertainties are commonly underestimated, we determined the uncertainties by identifying the standard deviation of sections of the light curve. For each object we identified 10 sections that contained slowly varying instrumental noise and spots, and Gaussian noise. We determined the noise of the 10 sections individually and subsequently used the average as our uncertainty value for all light curve points. To ensure we did not underestimate the noise in the light curve data, we included a Gaussian noise term in our fitting procedure, which is discussed in more detail in \S\ref{sec:gp}.

\subsubsection{The PHOEBE Model} \label{sec:ph_model}

The \ph\ modeling software combines the complete treatment of the Roche potential with the detailed treatment of surface and horizon effects such as limb darkening, reflection and gravity brightening to derive an accurate model of the binary parameters. The current implementation uses the WD method of summing over the discrete trapezoidal surface elements, which cover the distorted stellar surfaces, to determine an accurate representation of the total observed flux and, consequently, a complete set of stellar and orbital parameters. \ph\ incorporates all the functionality of the WD code, but also provides an intuitive graphical user interface alongside many other improvements, including updated filters, PHOENIX model atmospheres \citep{Husser2013} and bindings that enable interfacing between \ph\ and \python. 

To decrease the computational cost of using \ph\ with \emcee, for each iteration we created the light curve model containing 2000 data points in phase space. We then unfolded this model light curve into time space prior to adding the noise model and determining the log likelihood. This allowed the light curve to be computed in a relatively short amount of time, but fit the model of the light curve combined with the instrumental and stellar trends to the complete data set. 

The light curve and radial velocity data were fit simultaneously. Within our models, we fit the following parameters: inclination, $i$; eccentricity, $e$; argument of periastron, the angle from the ascending node to periastron, measured in the direction of motion, $\omega$; the primary and secondary potentials, proportional to the inverse radius, $\Omega_1$ and $\Omega_2$, respectively; third light $l3$; gamma velocity, the motion of the center of mass of the binary, $\gamma$; mass ratio, $q$; and semi-major axis, $sma$. We further set the albedos (reflective properties of the stars) and gravity darkening exponents (which relate to the change in temperature of the stars due to their obliquity), to the theoretical values of $A$\,=\,0.6 \citep{Rucinski1969b} and $\beta$\,=\,0.32 \citep{Lucy1967} for stars with convective envelopes ($T_\mathrm{eff} < 7000$\,K) and $A$\,=\,1.0 \citep{Rucinski1969a} and $\beta$\,=\,1.0 \citep{vonZeipel1924} for stars with radiative envelopes ($T_\mathrm{eff} > 7000$\,K).  As \citet{Diaz-Cordoves1992} showed that the square-root limb darkening model is preferable for objects that radiate towards the IR, we applied the square-root limb darkening law to our models and updated the limb-darkening coefficients after each iteration. Limb darkening is a thorny issue, with its implications discussed in detail in \citet{prsa2016}; it affects eclipse ingress and egress and is degenerate with stellar radii. We use limb darkening coefficients computed from the PHOENIX model atmospheres for the range of temperatures, surface gravities and chemical abundances applicable to our systems. We enforced consistency by interpolating the limb darkening coefficients for any explored combination of atmosphere parameters. While there are inherent limitations of using the square-root limb darkening model, the compounded systematic effects that arise from its use (compared to other limb darkening models) are below 0.5\%.

\subsubsection{Markov Chain Monte Carlo Derived Uncertainties\label{mcmcanalysis}}

Objective comparison between masses and radii measured observationally
and those predicted by theory requires that realistic uncertainties be
derived for the measurements. We use a Markov Chain Monte Carlo
(MCMC) sampler to sample the posterior probability distribution function. At each iteration we compute the posterior log-probability distribution function:

\begin{equation}
\log{P(\theta|D)} = \log{P(F|\theta)} + \log{P(RV|\theta)} + \log{P(\theta)} + C,
\end{equation}

where $D$ denotes the data, $F$ are the light curve measurements, $RV$ are the radial velocity measurements, $\theta$ is the parameter vector that contains the fitted parameters (specified in \S\ref{sec:ph_model}) and $C$ is an arbitrary constant. We incorporate \emcee\ into our analysis to sample the probability distribution function within a Bayesian framework. A significant advantage of this is that the results are presented as posterior probability distribution functions, which indicate parameter correlations and provide more robust uncertainty estimates. 

\mcmc\ explores the parameter space using a set of walkers, in our case 142. These chains begin with a uniform prior probability distribution for each parameter. At every iteration, each chain assesses its likelihood with respect to that of another chain and then elects whether to move towards that chain. The new parameters are based on the covariance matrix of the two chains. If the move increases the posterior likelihood then it is accepted; if the move decreases it then it may be accepted with a certain probability. During the initial burn-in time, Markov chains are converging towards their maximum likelihood position. The statistics of a large number of iterations provide converged posterior distributions for the model parameters. 

We create a model of all available data (light curves and radial velocities) for each binary star. To assess convergence, we use auto-correlation timescales. The auto-correlation timescale is used to estimate the number of iterations required to generate an independent sample, i.e.~the number of iterations required for the chain to ``forget'' where it started. We require a minimum of 30 auto-correlation time-scales to achieve convergence.

Eclipse depths are related to the temperature ratio. To obtain an accurate uncertainty for the temperature ratios, and thus secondary component temperatures, we marginalized over the primary and secondary effective temperatures, and report the posterior probability of the temperature ratio. From this, we calculate the secondary temperature and its uncertainty. The minimal rotational radial velocity that can be derived from APOGEE spectra is limited by the resolving power, allowing us to measure $v_\mathrm{rot} \gtrsim 10$\,km/s. We do not detect any broadening for any component in the studied systems, implying that all measured rotational velocities are below the $10$\,km/s detection threshold.

\subsubsection{\bf{Gaussian Process Regression}}
\label{sec:gp}

A Gaussian process (GP) is used to model noise, both instrumental and astrophysical; it is defined as a collection of random variables for which any finite number have a Gaussian distribution with a specified covariance structure. When using GPs, we use the data to condition the GP prior so that the GPs are normally distributed with respect to the data. A significant advantage of GP modeling is that it handles correlations in the data that are poorly understood by specifying only the high-level properties of a covariance kernel. We elected to use GPs to address the issues associated with stellar variations due to spots, correlated noise and instrumental systematics that are present in the light curves for both objects. We further incorporated a white noise kernel to assess our computed uncertainties.

Prior to the application of Gaussian processes, we create a combined binary star light and radial velocity curve model using \ph\ and \emcee. We then apply Gaussian processes to the light curve model, which removes the need to associate the systematics in the light curve with an explicit functional form. That way, we are able to model a wide range of systematics with a small number of tunable parameters.

The kernel, or covariance function, describes the similarity between two adjacent data points. For our kernel, we elected to use a term that approximates a Matern 3/2 function, which has a slowly and a rapidly varying component: 

\begin{equation}
\label{eqn:matern}
k(\tau) = \sigma^2 \left[ (1+1/\epsilon)e^{-(1-\epsilon)\sqrt{3}\tau/\rho} (1+1/\epsilon)e^{-(1+\epsilon)\sqrt{3}\tau/\rho}\right]
\end{equation}

Parameter $\epsilon$ controls the quality of the approximation since, in the limit as ${\epsilon \to 0}$, this becomes the Matern-3/2 function. For our computations, we set $\epsilon$ to 0.01. The Matern 3/2 function was selected as it is capable of modeling the noise, stellar variations and instrumental systematics in the light curve using only two parameters, $\sigma$ and $\rho$. 

We additionally incorporated a white noise term into our model. The purpose of this term is to ensure that our uncertainties are well estimated. Poorly estimated uncertainties lead to higher values for the jitter term (high in relation to the associated noise level) and the addition of white noise of the form: 

\begin{equation}
\label{eqn:noise}
k(\tau_{n,m}) = \sigma^2\delta_{n,m}, 
\end{equation}

where $\sigma$ is a tunable parameter.

\section{Masses and Radii for two Example \Kepler{} EBs\label{solvedebs}}

Here we describe the results of KIC 2445134 and KIC 3003991 to demonstrate the pipeline introduced in \S\ref{analysis}. These objects were selected as they are total eclipsers, have small \q{}, and the data are adequate to derive precise masses and radii.  \added{We also include the effective temperatures and metallicities as derived from the SDSS DR14 \citep{Abo2018} ASPCAP \citep{Gar2016} pipeline.  Although ASPCAP is optimized for giant and subgiant spectra, independent tests \citep{Wil2018, And2019} have shown offsets between ASPCAP temperatures and metallicities for FGK dwarfs are typically well within the ASPCAP uncertainties for our targets.  Binary spectra if treated as single-component can result in significant systematic bias in ASPCAP parameters \citep{Elb2018}, but the effects are minimal below mass ratios of $q \sim 0.4$, where our two systems lie.}

\subsection{KIC 2445134}

\begin{figure}
\begin{center}
\includegraphics*[width=5.2in,angle=0]{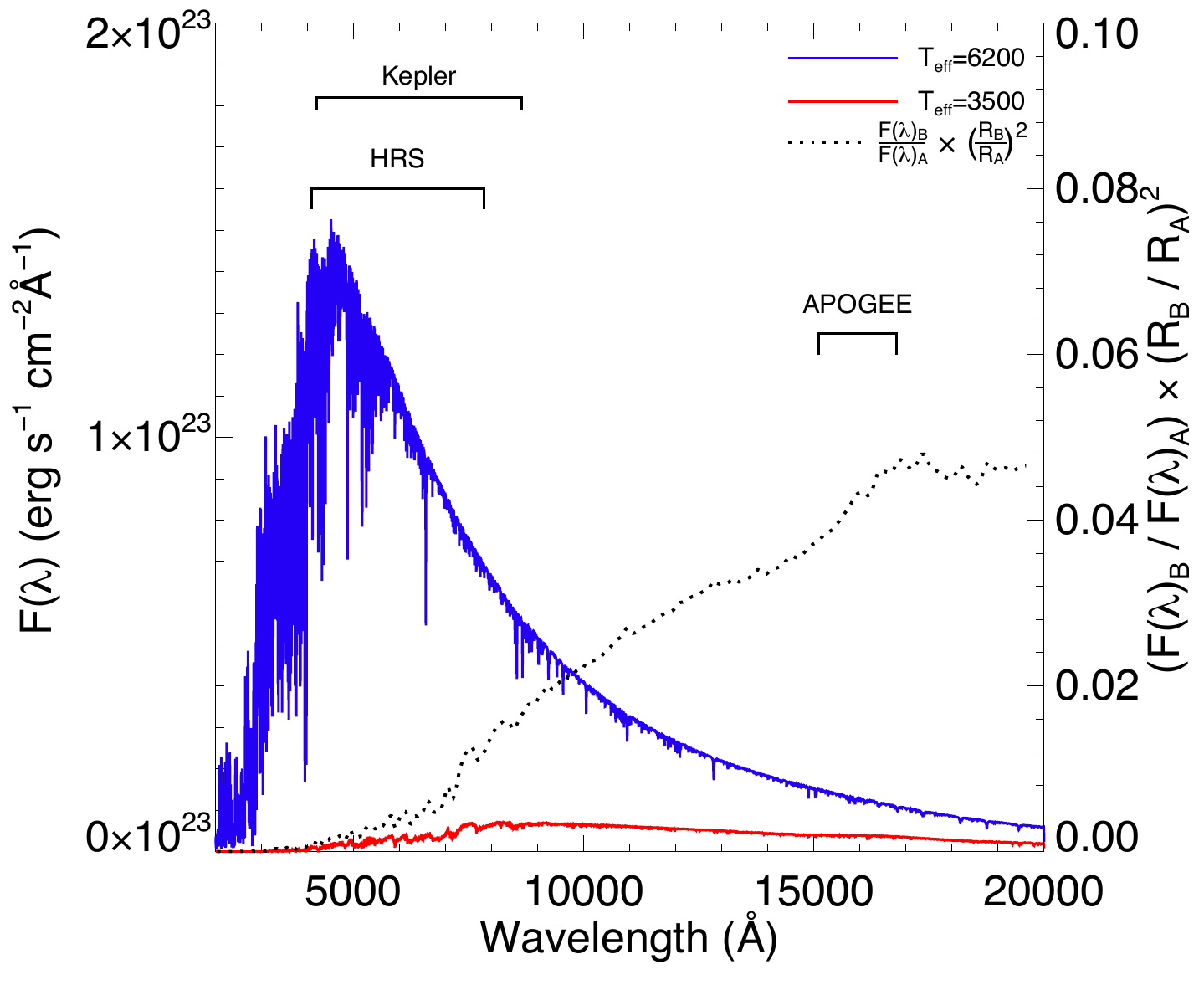}
\end{center}
\caption{BT-Settl models corresponding to the components in KIC 2445134
  (solid blue and red lines), and the flux ratio (dashed line).  The
  \Kepler{}, HRS, and APOGEE bandpasses are indicated for reference.
  The model spectra have been degraded to a spectral resolution of
  R=2,000, while the flux ratio is shown for R=100.  The contrast is
  an order of magnitude more favorable for detecting the secondary
  in the H-band than in the optical.} \label{fig:fluxratio}
\end{figure}

\begin{figure}
\begin{center}
\includegraphics*[width=5.2in,angle=0]{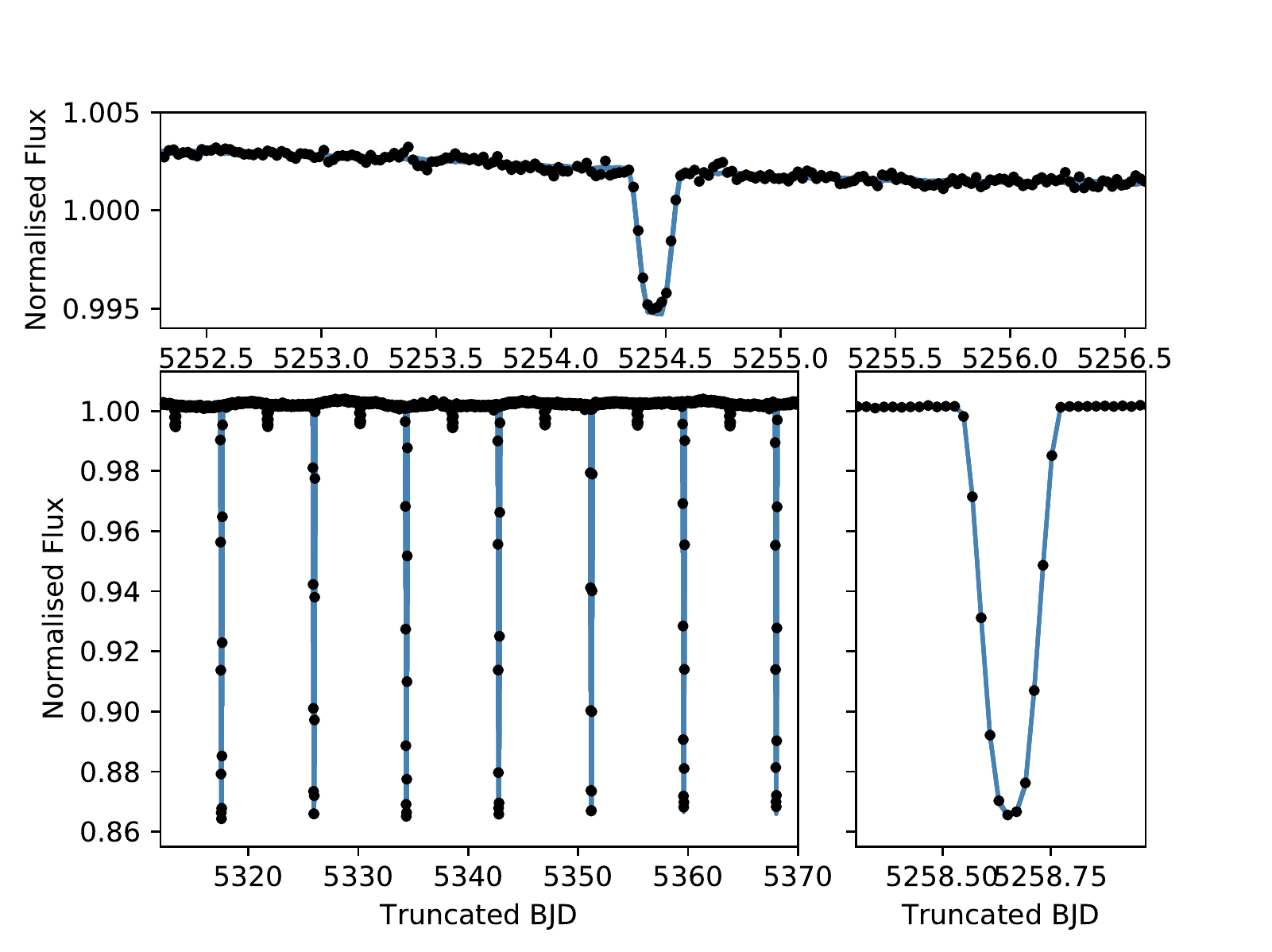}
\end{center}
\caption{\kep\ light curve (black) and ten models including GPs (blue) for KIC\,2445134, where the different panels emphasize different regions of the light curve.} \label{fig:model2445lc}
\end{figure}

\begin{figure}
\begin{center}
\includegraphics*[width=5.2in,angle=0]{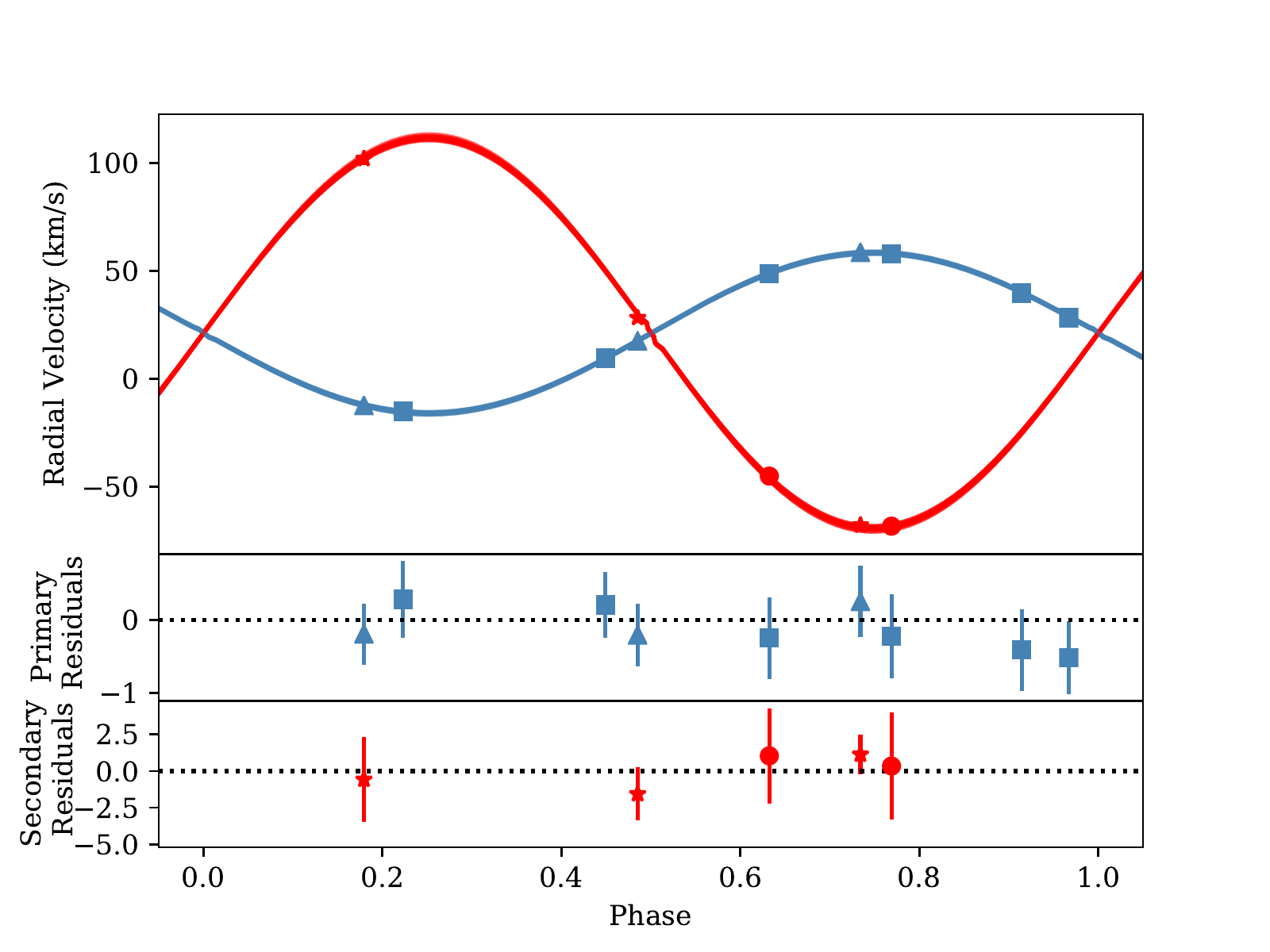}
\end{center}
\caption{Phase folded RVs for KIC\,2445134 derived from the HET spectra, squares and circles, and the APOGEE spectra, triangels and stars, for the primary and secondary components, respectively. The (\emph{upper panel}) depicts the radial velocities and ten models from the final iteration (where each model is depicted by a red or blue line). The (\emph{middle panel}) and (\emph{lower panel}) depict the residuals from the average model fit to the primary and secondary radial velocity data, respectively. The residuals are measured in $\kms$.} \label{fig:model2445rv}
\end{figure}

KIC 2445134 is comprised of an F-type dwarf primary and an M-type
secondary in an 8.4 day period, and has a flux ratio of $\sim0.01$ in
the \Kepler{} bandpass. It has a \Kepler{} magnitude of 13.55 and an
H-band magnitude of 12.40. \added{The ASPCAP $T_{\rm{eff}}$ is $6260 \pm 170$ K and the metallicity is [M/H] = $-0.046 \pm 0.105$.} We obtained six observations of KIC 2445134
with the HRS, which we solved as an SB1 using a mid-F type HRS
template, and three observations with APOGEE that we solved as an SB2
using the BT-Settl templates. These measurements constrained the
spectroscopic orbital parameters and allowed us to
re-examine our HRS cross-correlation functions and detect the faint
companion in two epochs using a mid-M star HRS template and with
correlation power slightly above the noise.  Table~\ref{rv2445134}
lists the date and corresponding barycentric Julian Date for each of
our observations, the RVs we measure, and the associated spectrograph.
Figure~\ref{fig:fluxratio} shows the BT-Settl models corresponding to
the primary and secondary components, along with the flux ratio for a
synthetic EB constructed from these models.  We carried out the MCMC
analyses described in \S\ref{mcmcanalysis} on the RVs in
Table~\ref{rv2445134} and the \Kepler{} photometry, and derived the
orbital and physical parameters listed in Table~\ref{orbpar2445134}.
Figures~\ref{fig:model2445lc} and\,\ref{fig:model2445rv} depict the light and radial velocity curve models, respectively. Due to the large mass ratio, temperature ratio and the orientation of the orbit, the light curve of KIC\,2445134 contains Doppler boosting. We incorporated Doppler boosting into our model using the framework of \citet{Bloemen2011} and present the Doppler boosting parameters for the primary and secondary components, $B_A$ and $B_B$, respectively, in Table~\ref{orbpar2445134}. Our measurement of the mass ratio, $\q=0.411\pm0.001$, has a precision of better than 1\%. Our derived masses ($\MA=1.29\pm0.03\,\msun$,
$\MB=0.53\pm0.01\,\msun$) and radii ($\RA=1.42\pm0.01\,\rsun$,
$\RA=0.510\pm0.004\,\rsun$) have measurement precisions of 1\% or
better, except for the primary mass which is constrained to 2\%. 

To assess the dependency of the parameter uncertainties on our light-curve per-point uncertainties, we ran our software in the same manner as outlined above, but with the uncertainties multiplied by 0.5 and 2 for KIC\,2445134. The outcome was that all fundamental parameters remained within the one sigma uncertainties quoted and the uncertainties were unchanged with the exception of the secondary radius uncertainty, which changed from 0.004\,$\rsun$ to 0.005\,$\rsun$ (still providing a 1\% uncertainty) for the case of the increased per-point uncertainties.

\subsection{KIC 3003991}
\begin{figure}
\begin{center}
\includegraphics*[width=5.2in,angle=0]{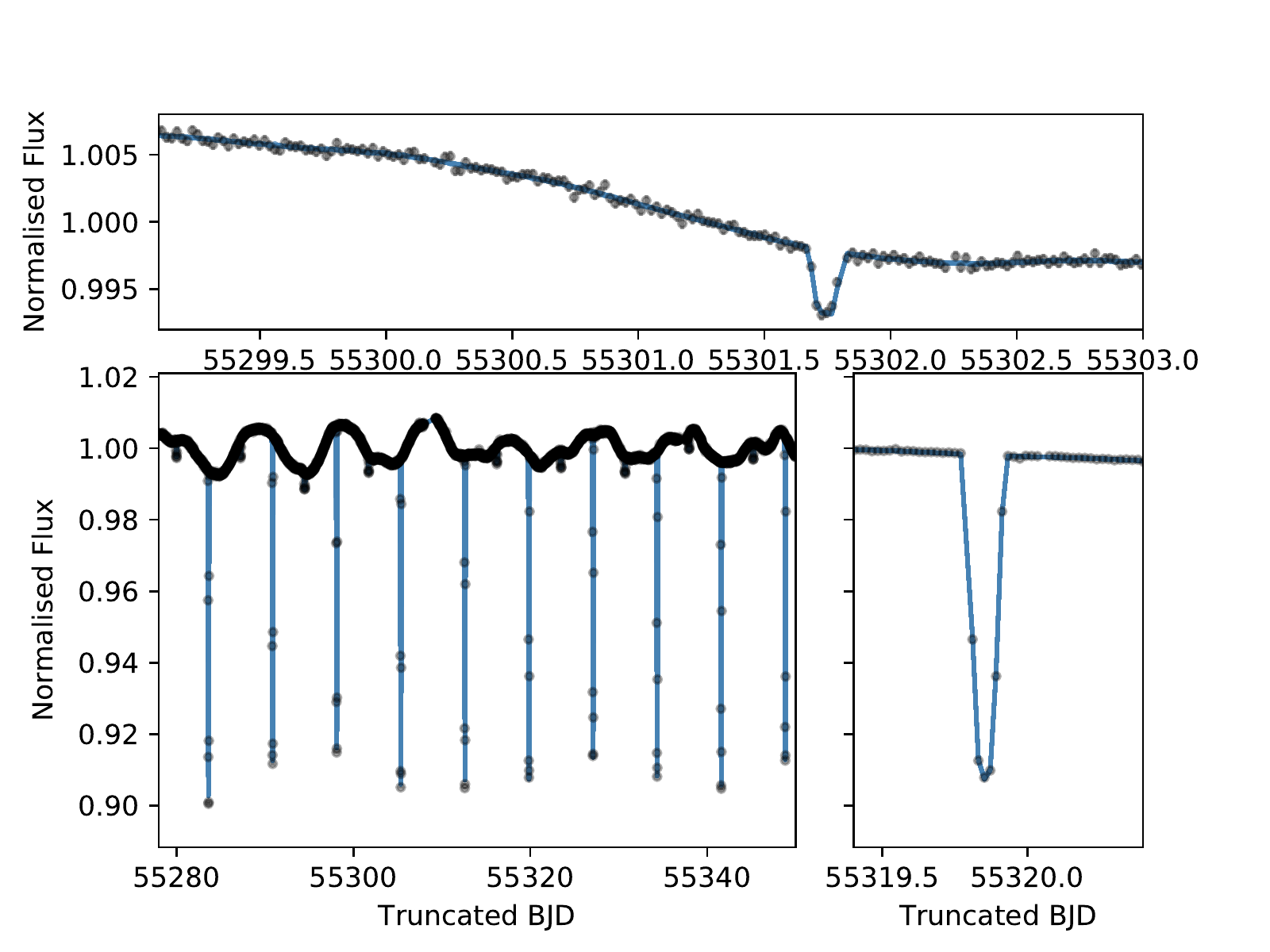}
\end{center}
\caption{\kep\ Kepler light curve (black) and ten models including GPs (blue) for KIC\,3003991, where the different panels emphasize different regions of the light curve.} \label{fig:model3003lc}
\end{figure}

\begin{figure}
\begin{center}
\includegraphics*[width=5.2in,angle=0]{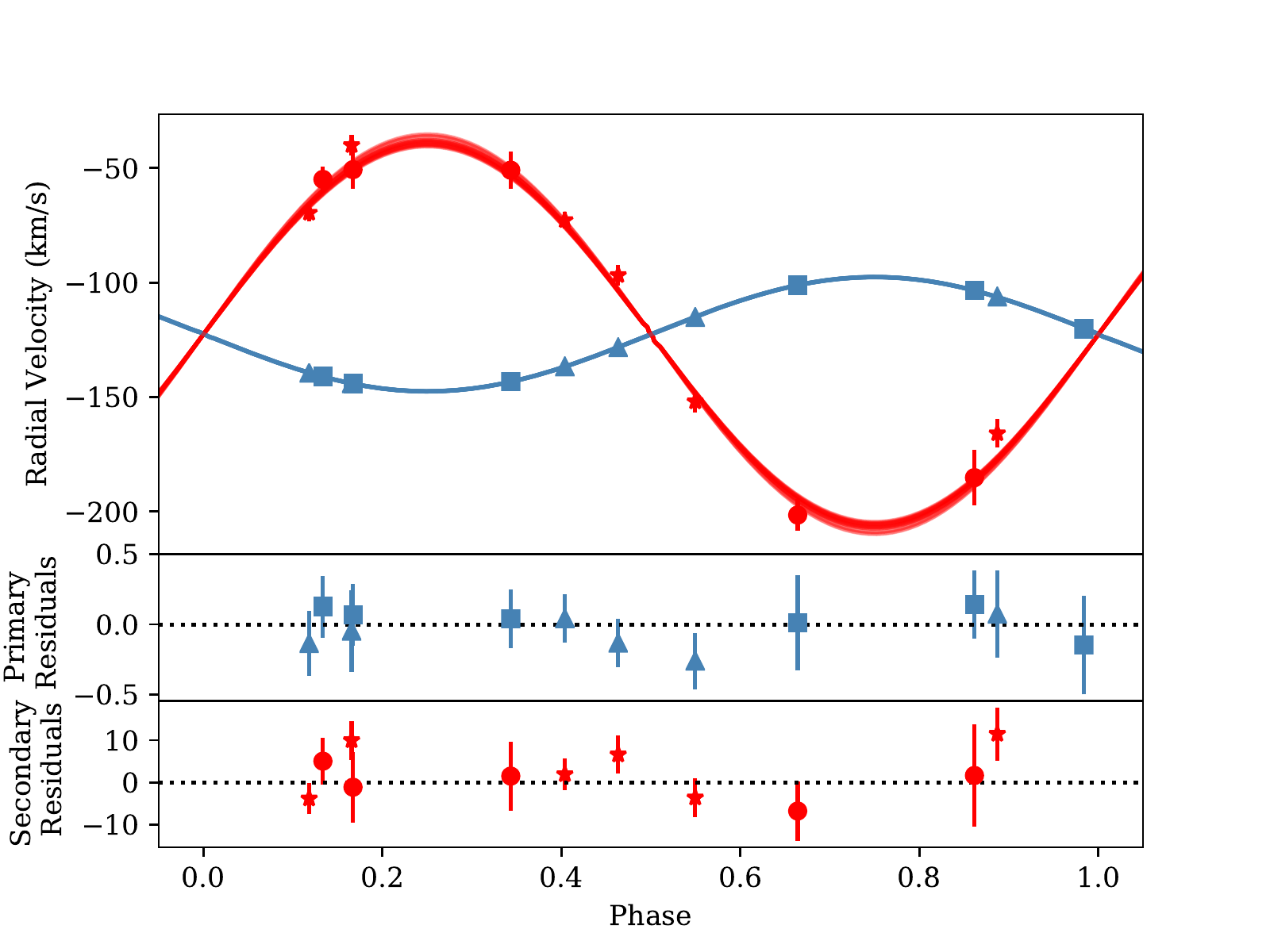}
\end{center}
\caption{Phase folded RVs for KIC\,3003991. The layout is identical to Figure\,\ref{fig:model2445rv}. The thickness of the lines depict the spread of the ten models caused by the model uncertainty. This is particularly significant for the secondary component (red lines).} \label{fig:model3003rv}
\end{figure}

\begin{figure}
\begin{center}
\centering
\hspace{-1.4cm}
\vspace{-2cm}
\includegraphics*[width=7.in,angle=0]{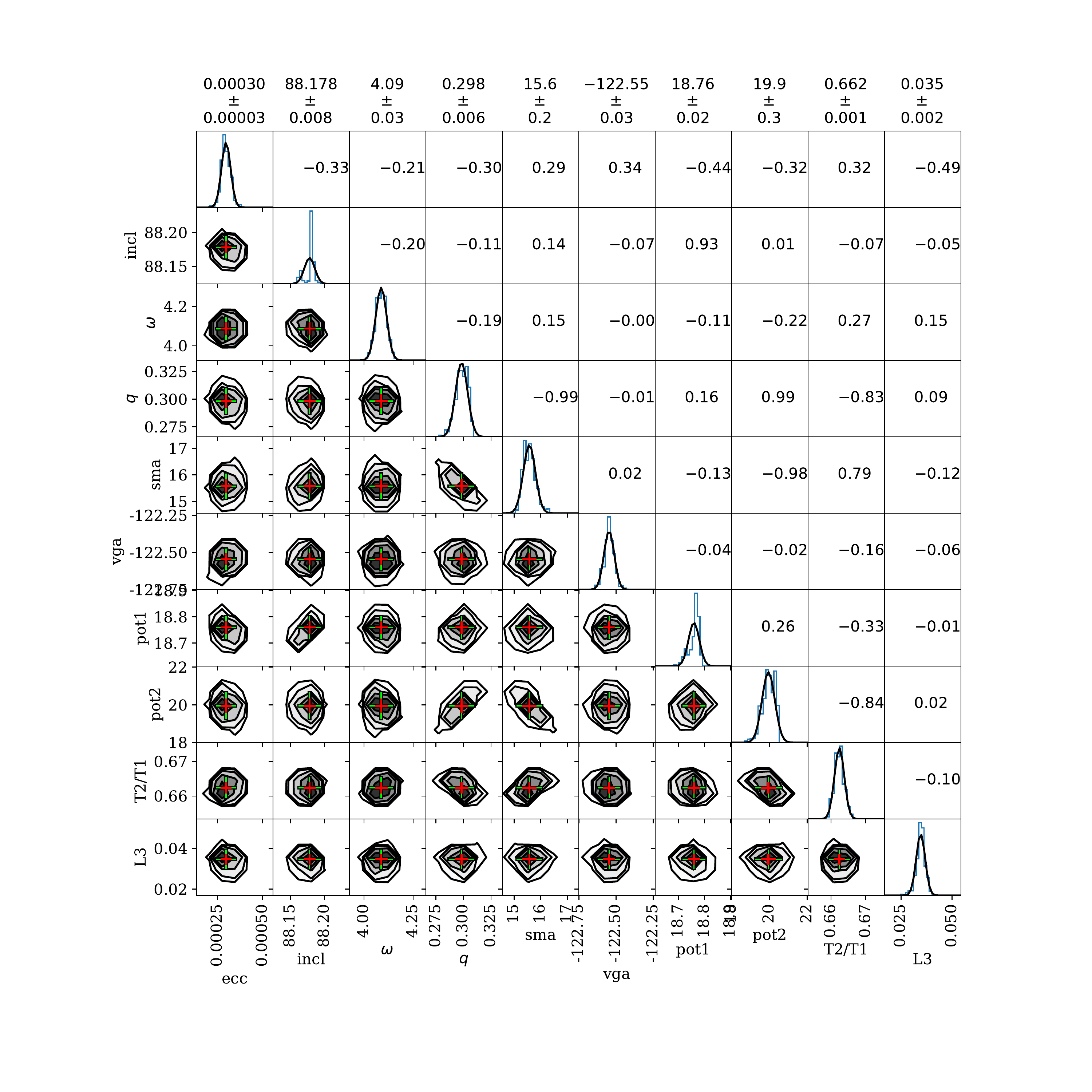}
\end{center}
\caption{Posterior distributions for the KIC\,3003991 parameters derived from \Kepler{} photometry, and APOGEE and HRS radial velocities. Boxes in the upper right contain the standard correlation coefficient, with 1 corresponding to perfect
  correlation, -1 perfect anti-correlation, and 0 no correlation.
  Final parameter values are listed along the top.} \label{fig:3003post}
\end{figure}

\begin{figure}
\begin{center}
\vspace{-2.cm}
\includegraphics*[width=6.5in,angle=0]{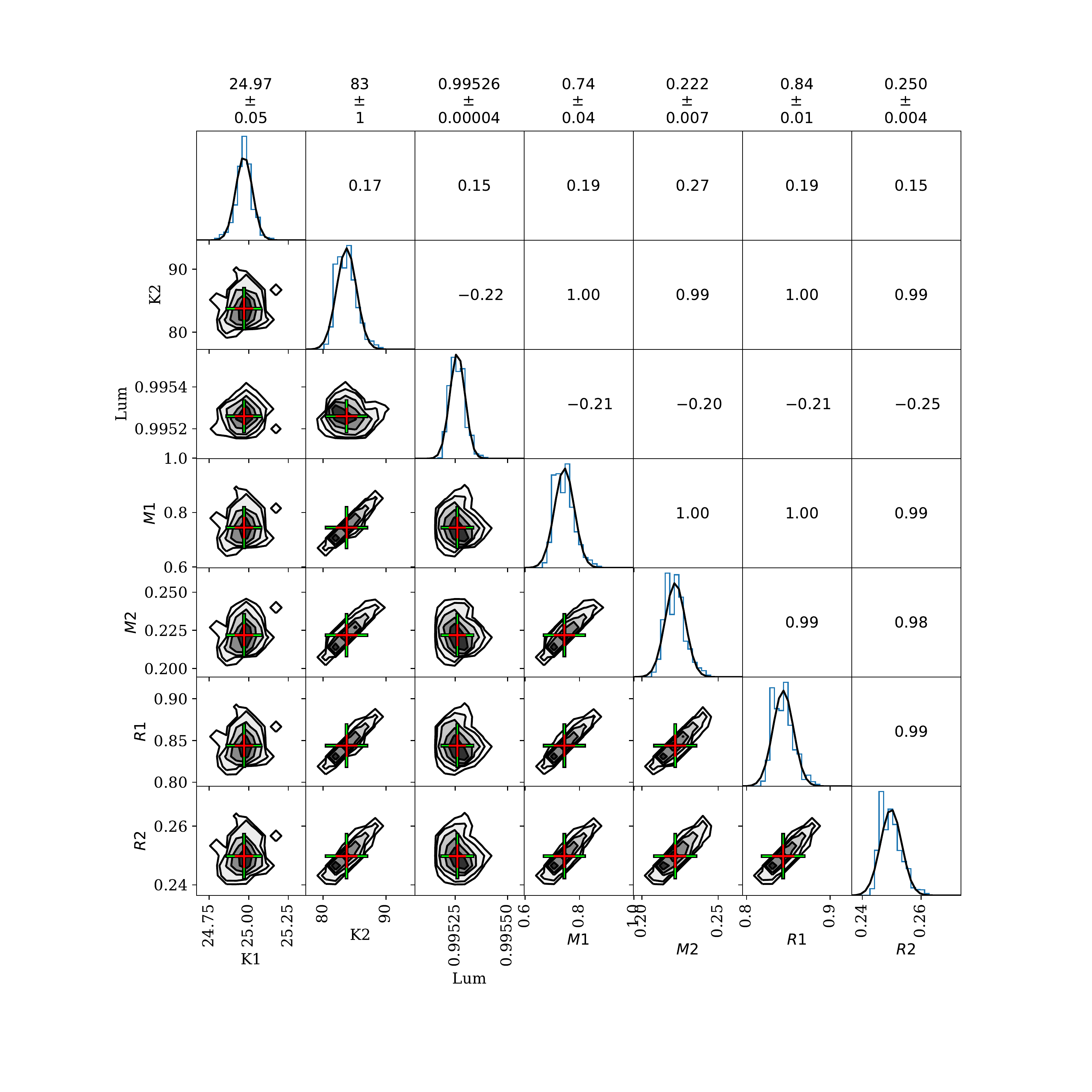}
\end{center}
\caption{Parameter distributions for the calculated parameters of KIC\,3003991. Layout is as Figure\,\ref{fig:3003post}} \label{fig:3003postblob}
\end{figure}

KIC 3003991 represents the faint end of our EB sample, with \Kepler{}
mag = 13.9, and is comprised of a late G-star and a mid M-star in a
7.2 day orbit with a \Kepler{} flux ratio of $\sim$0.005.  \added{The ASPCAP $T_{\rm{eff}}$ is $5340 \pm 140$ K and the metallicity is [M/H] = $-0.41 \pm 0.073$.}  We
obtained six observations with HRS and six observations with
APOGEE. Once again, we were able to easily detect the low mass
companion in the H-band spectra, with $\alpha\sim0.10$, and used the
measurement of \KB{} from the APOGEE spectra to re-analyze the HRS
spectra and recover the companion RV for many of the HRS epochs.  The
relative faintness of this EB resulted in a poorer S/N than KIC
2445134 for both the APOGEE and HRS spectroscopy, which manifests as
larger uncertainties on primary and secondary RVs.
Table~\ref{rv3003991} lists the date and corresponding barycentric
Julian Date for each of our observations, the RVs we measure, and the
associated spectrograph.  Figure~\ref{fig:model3003lc} and Figure~\ref{fig:model3003rv} depict the light curve model and phase
folded spectroscopic orbit for KIC 3003991, corresponding to the orbital and physical parameters in Table~\ref{orbpar3003991}. Figure~\ref{fig:3003post} shows the posterior
distributions and Figure~\ref{fig:3003postblob} shows distributions of quantities calculated from our model posteriors (where the calculations were performed after each iteration).  We measure a mass ratio of $\q=0.289\pm0.006$, and derive masses of $\MA=0.74\pm0.04\,\msun$ and $\MB=0.222\pm0.007\,\msun$, corresponding to precisions of 5\% and 3\%, respectively.  The uncertainty on both is driven largely by the uncertainty on the semi-amplitude, which
is difficult to reduce further without a significant additional
spectroscopic investment.  As the precision we derive for \MB{}
complies with our program objective of 3\%, it will provide
useful constraints on stellar models due to the paucity of
precisely measured masses in the $M\la0.25\,\msun$ regime.
Additionally, the radii we measure for this EB,
$\RA=0.84\pm0.01\,\rsun$ and $\RB=0.250\pm0.004\,\rsun$, have
precisions of 1\% and 2\%, respectively.

\section{False Positives in our Sample}
\begin{figure}
\begin{center}
\includegraphics*[width=5.2in,angle=0]{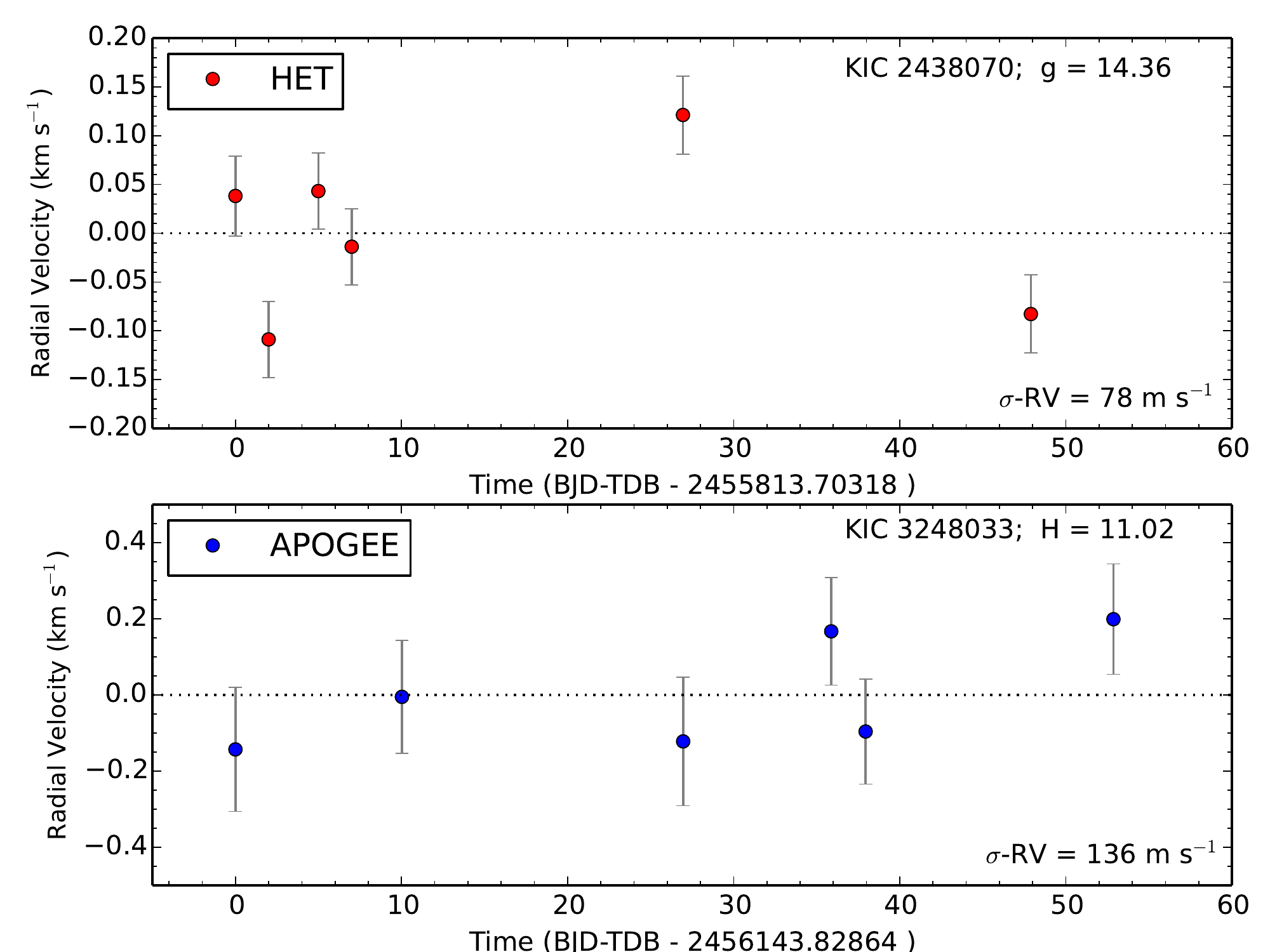}
\end{center}
\caption{RV measurements of KIC 3248033 identified as a false-positive based on
  pixel level analysis of the photometry and of a red-giant KIC 2438070 with no stellar eclipsing companion. Our RV measurements
  dynamically verify that the eclipse signal in the light-curve data results from an unresolved background EB which is highly diluted by
  the foreground KIC target.} \label{fig:falsepos}
\end{figure}

We selected our EB sample (Table~\ref{ebsampletable}) in 2011, while
the analysis pipeline being used by the \Kepler{} team to identify EBs
and substellar companions was still being refined.  In a few cases,
the eclipse signal detected in \Kepler{} photometry results from a
background EB that is spatially unresolved from the primary KIC target
at the \Kepler{} plate scale \citep{Abdul2016}.  Many of these ``false-positives'' have
subsequently been identified by the \Kepler{} team using advanced
light-curve analysis of the photometry at the pixel level, and have
been flagged in the \emph{Kepler Eclipsing Binary Catalog}.  We
included 14 stars now labeled as false-positives for which we have
APOGEE or HRS spectra to look for RV variability.  In all cases, these
targets show no variability within the RV measurement precisions
listed in \S\ref{facilities}.  These targets, which we have
dynamically verified as false positives, are identified in the last
column of Table~\ref{ebsampletable}.  Figure~\ref{fig:falsepos}
illustrates the attained RV precision of a star (top panel) that does not have
a stellar component with a period within our baseline, amounting to better than $\sim$100\,m/s. The magnitude
and RV precision of this star is representative of stars observed from
this program with the HET. The RV measurement precision of a false
positive observed with APOGEE is also shown (bottom panel), and is
representative of the typical RV uncertainty for our APOGEE targets.

\section{Discussion \& Future Prospects}
\begin{figure}
\begin{center}
\includegraphics*[width=5.2in, height=4.8in,angle=0]{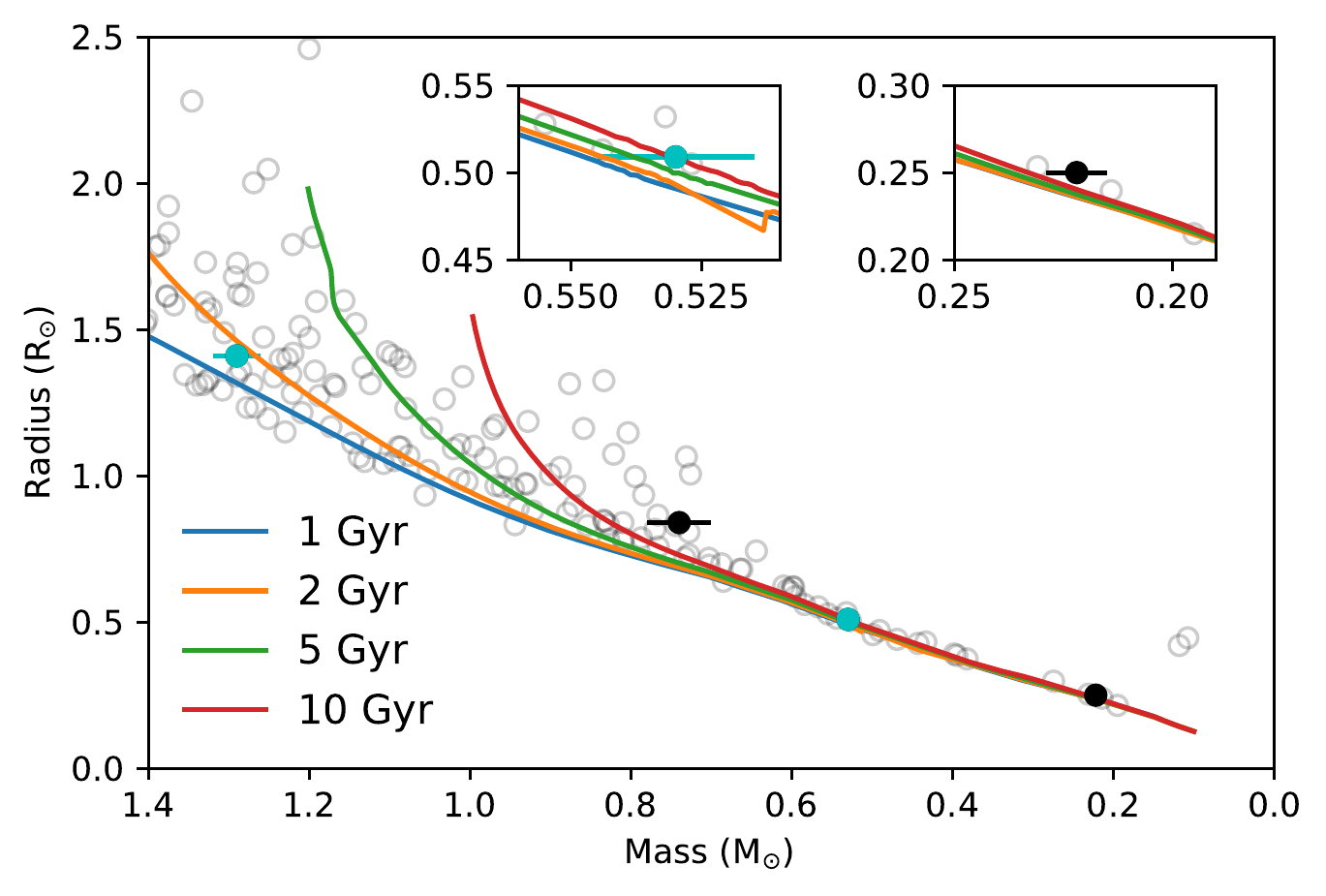}
\end{center}
\caption{Radius versus mass, plotted for the EBs in the DEBCat
  Eclipsing Binary Catalog (open circles), along with the two EBs we
  describe in \S\ref{solvedebs}, where KIC\,2445134 is depicted with solid blue circles and KIC\,3003991 is depicted with solid black circles.  For comparison, we show {\sc{mist}} ({\sc{mesa}} Isochrones and Stellar Tracks) stellar isochrones (\citealt{Dotter2016,Choi2016}) for solar metallicity stars. The inserts depict magnified regions for the secondary components of KIC\,2445134 (left insert) and KIC\,3003991 (right insert). The paucity of precision measurements below
  $\sim0.8\,\msun$, and particularly below $\sim0.25\,\msun$, is
  evident.} \label{fig:debcatcomp1}
\end{figure}

\begin{figure}
\begin{center}
\includegraphics*[width=5.2in, height=4.8in,angle=0]{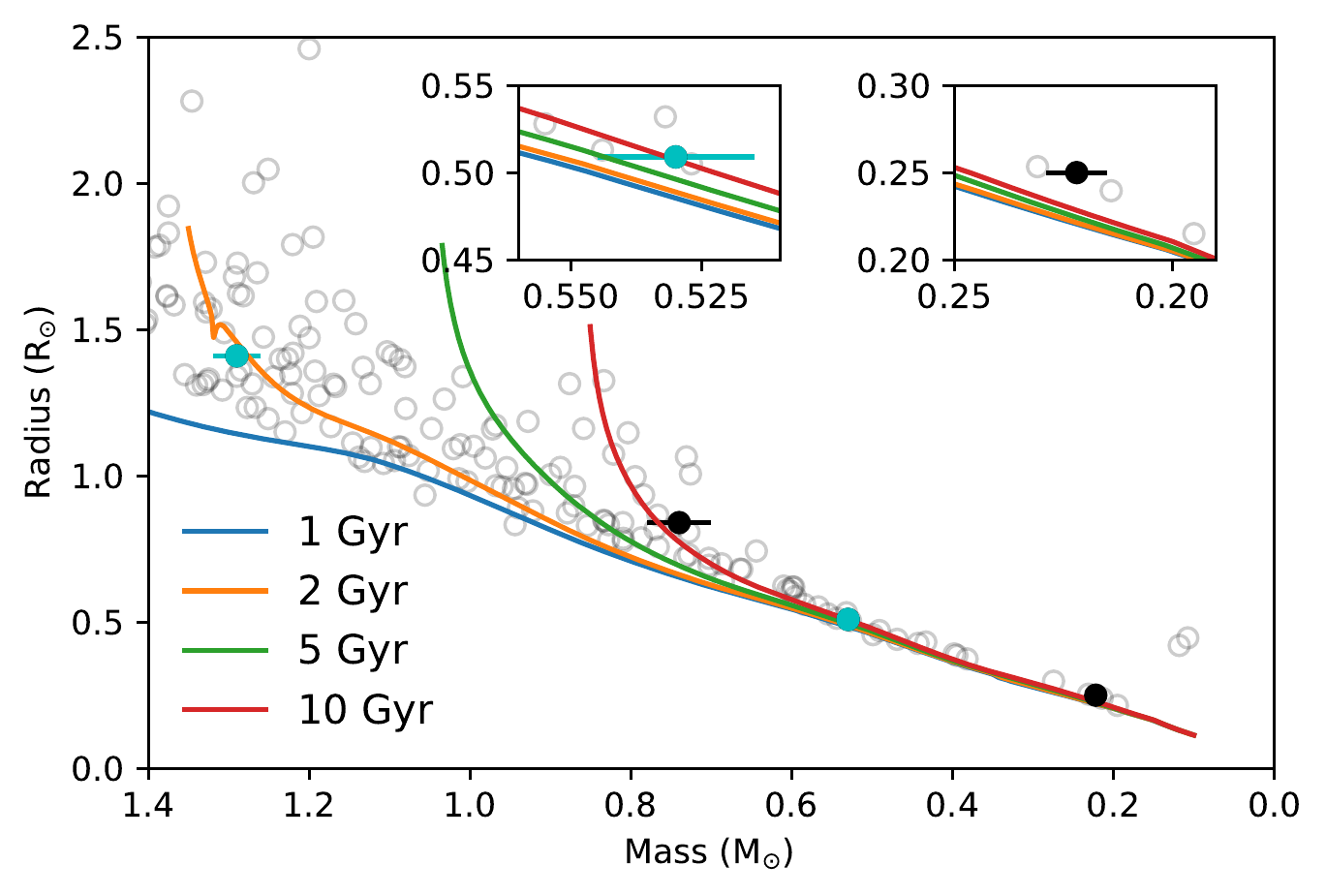}
\end{center}
\caption{The same as Figure\,\ref{fig:debcatcomp1}, but with [Fe/H]=-1. In this case, the primary component of KIC\,3003991 now agrees with the 10\,Gyr isochrone, however, the secondary component does not. This is inline with a known disparity between theoretical stellar models and observations in the low mass region, and further highlights the importance of modeling low mass binary components \citep{2010A&ARv..18...67T}.} \label{fig:debcatcomp2}
\end{figure}

The two EBs presented in \S\ref{solvedebs} illustrate the
contributions made by our ongoing work to the direct and precise measurement of
masses and radii for low-mass stars. Many systems in our EB sample have
primary components with mass $M > 0.8\,M_{\odot}$, a regime where
models of stellar mass and radii are well matched to observational
constraints. Many have small mass ratios ($q<0.6$), and so have
low-mass secondaries, specifically targeting the regime with
disagreement between models and existing observations. The high mass
primary allows traditional spectroscopic analysis tools to obtain
metallicity for the spectra since the secondary star flux is low
enough that the effect of spectral contamination is minimal. As
discussed in \citet{ 2012ApJ...760L...9T} the stellar metallicity is
an important constraint on the models, as well as being necessary to
derive ages. This project aims to derive masses
and radii for low mass stars using a combination of precision
photometry from {\it Kepler} coupled with radial velocities from
stable fiber-fed optical and near-infrared high resolution
spectrographs.  Figure \ref{fig:debcatcomp1} depicts the mass-radius
relationship for the stars in this work (filled blue and black circles), compared with those from DEBCat (open circles). Figure \ref{fig:debcatcomp2} depicts the same but with [Fe/H] = -1 isochrones (as opposed to Solar metlicity). The primary and secondary component of KIC\,2445134 both show good agreement (within 1$\sigma$) with the 2\,Gyr Solar metalicity isochrones. The primary component of KIC\,3003991 shows good agreement (within 1$\sigma$) with the 10\,Gyr isochrones, however, the secondary component is 3$\sigma$ away. This kind of disagreement is not unusual in the literature for low mass stars \citep{2010A&ARv..18...67T}. Our program with APOGEE will
continue to fill in parameter space by obtaining double-line
spectroscopic RVs for high-flux-contrast systems. KIC 2445134 and KIC 3003991 would have been especially challenging without
the benefit of NIR spectroscopy. For these systems we succeeded in obtaining the 3\% program goal for the uncertainties on the masses and radii (with the exception of the primary mass of KIC\,3003991), making these valuable benchmark
systems. Additional results emerging from this program will further
populate the mass-radius plane for low mass dwarfs, study systems with
equal-mass components, and provide high precision checks on parameters
derived from asteroseismology for a few, rare giants in EB systems
with detected stellar oscillations.

\acknowledgments We acknowledge support from two NASA ADAP grants (NNX13AF32G and 16-ADAP16-0201), and NSF grant AST 1517592 for this project. This work was partially supported by funding from the Center for Exoplanets and Habitable Worlds. The Center for Exoplanets and Habitable Worlds is supported by the Pennsylvania State University, the Eberly College of Science, and the Pennsylvania Space Grant Consortium. We acknowledge support from NSF grant AST 1006676 and AST 1126413 in our pursuit of precision radial velocities in the NIR. This research has made use of the SIMBAD database, operated at CDS, Strasbourg, France. This publication makes use of data products from the Two Micron All Sky Survey, which is a joint project of the University of Massachusetts and the Infrared Processing and Analysis Center/California Institute of Technology, funded by the National Aeronautics and Space Administration and the National Science Foundation. Finally, we acknowledge the thorough review of the anonymous reviewer whose comments saved us from a small embarrassment.

This work was based on observations with the SDSS 2.5-meter telescope. Funding for SDSS-III has been provided by the Alfred P.~Sloan Foundation, the Participating Institutions, the National Science Foundation, and the U.S. Department of Energy of Science. The SDSS-III web site is http://www.sdss3.org/. SDSS-III is managed by the Astrophysical Research Consortium for the Participating Institutions of the SDSS-III Collaboration including the University of Arizona, the Brazilian Participation Group, Brookhaven National Laboratory, University of Cambridge, Carnegie Mellon University, University of Florida, the French Participation Group, the German Participation Group, Harvard University, the Instituto de Astrof\'isica de Canarias, the Michigan State/Notre Dame/JINA Participation Group, Johns Hopkins University, Lawrence Berkeley National Laboratory, Max Planck Institute for Astrophysics, Max Planck Institute for Extraterrestrial Physics, New Mexico State University, New York University, Ohio State University, Pennsylvania State University, University of Portsmouth, Princeton University, the Spanish Participation Group, University of Tokyo, University of Utah, Vanderbilt University, University of Virginia, University of Washington, and Yale University.

Data presented herein were also obtained at the Hobby Eberly Telescope (HET), a joint project of the University of Texas at Austin, the Pennsylvania State University, Stanford University, Ludwig-Maximilians- Universita\"at Mu\"nchen, and Georg-August-Universit\"at G\"ottingen. The HET is named in honor of its principal benefactors, William P. Hobby and Robert E. Eberly.

\appendix
\section*{Appendix A - Detailed Description of Our HET HRS Pipeline}
\label{appendixa}
Our EB project makes use of an internal data reduction pipeline for the 
HRS on HET.  For future reference with other papers in this series, and for
others using this instrument on the HET,  we provide a
detailed description of our pipeline, which is composed of a series of
independent subroutines that can be combined into nightly processing
scripts; we refer to on-sky observations of science targets
generically as target frames, and individual calibrations either
generically as calibrations or by their specific function.  As an
aside, the pipeline can easily process HRS data taken in any of the
possible HRS configurations, and so, as a service to other observers
who may be interested in utilizing its functionality, the description
we provide here is intended to be general, rather than specific to the
HRS configuration listed above.

A raw HRS FITS image contains three components: the main FITS header
describes the instrument configuration and target exposure, while the
first and second FITS extensions contain the red and blue image data,
respectively.  Instrument keywords in the main header can reliably
reconstruct the HRS configuration used for any given observation,
without relying on external log sheets or records.  Keywords
describing the target, such as OBSTYPE and OBJECT, are manually
specified by the HET resident astronomer carrying out the observation,
and are not always used consistently.  To begin reducing a night's
observations, we copy the raw target and calibration images into a
single working directory.  The pipeline automatically classifies the
various calibration and target frames, and carries out basic image
processing on them.  If the working directory contains images taken
with different HRS configurations, each configuration will be parsed
and handled automatically.

First, \texttt{HRS\_SPLIT} extracts the red and blue images from the
multi-extension FITS file, saves them as individual zero-extension
FITS images with a suffix of '\_r' or '\_b', and assigns the data
contained in the original header.  This step ensures that the original
FITS files are not modified in any way by the pipeline.  We then
perform an overscan correction (\texttt{HRS\_OVERSCAN}) by triming
each image to the data region and subtracting a pedestal value
determined by computing the median of the overscan region.  Higher
order corrections that could be derived from the overscan are instead
corrected with the bias frames.  Different HRS configurations utilize
different on-chip CCD binning schemes, and \texttt{HRS\_BIAS}
automatically groups and combines bias frames based on their binning,
producing master bias frames for each configuration used during a
night.  A similar process is then performed on all available flat
fields (\texttt{HRS\_FLAT}), although the number of possible
configurations is much larger for flats than for bias frames.
\texttt{HRS\_FLAT} properly separates multiple sets of flats from the
same configuration that were obtained with different exposure times,
which is occasionally necessary when very high signal-to-noise flats
are required.  In addition, non-standard flat fields taken through the
HRS target fibers illuminated by the Medium Resolution Spectrograph
\citep{2003SPIE.4841.1036R} flat lamp or twilight sky are properly
recognized.  Each master flat field is automatically bias corrected
and saved with a unique name identifying the light source, instrument
configuration, and exposure time.  Similar corrections are applied to
individual target frames.  These image processing steps can generally
be carried out without any human input or supervision, and are
conveniently grouped in a script, \texttt{HRS\_IPROC}, which requires no
input parameters or keywords.

Extracting one dimensional spectra from a two dimensional image
requires locating and tracing each echelle order in the target
image.  The HRS dispersion runs vertically across the detector (our
\emph{y} coordinate), and we make the simplifying (but essentially
correct) assumption that the HRS slit is aligned perfectly with
detector pixel rows (our \emph{x} coordinate).  Each order is traced
independently, and each trace begins by identifying one \emph{x} \&
\emph{y} pair that intercepts the fiber profile.  At that position we
extract a subarray with an \emph{x} dimension of 1.5 times the
physical width of the fiber in pixels, which fully encloses the
illuminated pixels without overlapping into the adjacent orders, and a
\emph{y} dimension of 4 pixels, which averages over noise and
increases the speed of the tracing procedure.  The illumination
profile of a fiber resembles a top hat, with some variation across the
illuminated pixels due to imperfect radial scrambling in the fiber.
This functional form is not well represented by the Gaussian
profile commonly used for slit fed spectrographs.  Instead, to
identify the \emph{x} coordinate of the beam center we compute the
derivative of the profile and fit that with a combination of two
Gaussians of equal width, separated by the slit width in pixels.  This
parameterization reliably recovers the beam center position at the
selected \emph{y} coordinate.  The algorithm then walks up and down
the beam, computing \emph{x} for each binned \emph{y}.  We fit these
measured beam positions with a low order polynomial to derive the beam
position at each dispersion position.

The tracing algorithm is implemented in two separate subroutines:
\texttt{HRS\_MTRACE} and \texttt{HRS\_ATRACE}.  \texttt{HRS\_MTRACE}
allows the user to interactively identify beams, and annotate each
with the echelle order and the type of spectrum (absorption, such as a
target or flat field; or emission, such as a sky or ThAr). This
information is encapsulated in a template file that fully describes
the HRS configuration.  \texttt{HRS\_ATRACE} is fully automated, and
uses previously derived templates to process routine sets of images,
retracing all beams to compensate for any misalignment between the
template and the target frame due to imprecision in repositioning the
HRS components.

Preserving the relative flux information in our science target frames
requires that the master flat frames be normalized to remove both the
echelle blaze and the fiber illumination profile.
\texttt{HRS\_FLATNORM} derives this normalization, using a procedure
conceptually similar to optimal spectral extraction.  Such schemes
have been described extensively in the past
\citep[e.g.][]{1986PASP...98..609H, 1990PASP..102..183M,
  2004PASP..116..362C}, and we have adapted them for use with a
fiber-fed echelle spectrograph.  Each master flat is traced, and then
each beam is rectified.  The rectified beam is used to derive a
dispersion dependent spatial profile for the beam, following the
procedure outlined by \citet{2004PASP..116..362C} and figures therein.
This profile is the desired function for normalizing the flat fields.
Reversing the image rectification on the beam profile yields the
normalization image that can be applied to the original master flat
frame.

To extract the individual target spectra, \texttt{HRS\_OPTEXT} divides
the normalized flat field into the two dimensional target image, and
then applies an optimal extraction algorithm similar to that used for
normalizing the flat fields.  If multiple exposures of the same target
were obtained, the user has the option to extract each image
individually, or coadd them in two dimensions and extract the
composite spectrum.  Optimal fiber profiles are automatically derived
for each beam, and provide the proper weighting functions for optimal
extraction.  A primary benefit often associated with optimal
extraction algorithms is their ability to optimize the S/N of the
extracted spectrum by minimizing the contribution of detector noise.
While this is useful for some of our fainter \Kepler{} EB targets, more
useful is the algorithm's ability to automatically identify bad pixels
in the two dimensional images caused by detector defects or
cosmic-rays.  These pixels are automatically excluded and the
weighting function at each affected dispersion element is adjusted
accordingly.  \texttt{HRS\_OPTEXT} returns optimally extracted
spectra, sum extracted spectra for comparison purposes, and wavelength
dependent variances that provide realistic measures of the spectrum
S/N.  These data products, as well as descriptions of each echelle
order and beam, are encapsulated in a FITS table for each target
image.

Our wavelength calibration pipeline relies on a construct similar to
that behind the tracing algorithms: time intensive manual line
identification defines templates for each HRS configuration, and these
templates are used in an automated re-identification procedure for
routine data processing.  The templates are generated using the IDL
based WAVECAL package \citep{2002A&A...385.1095P} to associate the
wavelengths of known ThAr spectral lines with their \emph{y} pixel
location for each spectral order.  Wavelengths are taken from the
linelist of \citet{2007MNRAS.378..221M}, and typically number $\sim10-20$ per order.  The set
of lines comprising each order is fit with a low order polynomial
(typically 4th order) to derive the template dispersion solution for
that order.  To calibrate a target spectrum, we automatically measure
the ThAr image taken consecutively to the target observation, using
the template as a guide.  The dispersion solution for each order is
automatically solved for, using a sigma-clipping rejection to detect poorly fit
lines, and the resulting wavelength solution is applied to the corresponding
target spectrum.  Although the HRS is not a pressure or temperature
stabilized instrument, this careful use of ThAr calibrations
consecutive with target observations can yield radial velocity
stability of $25\,\mathrm{m s^{-1}}$ or better for observations taken
on different nights, weeks, or even months
\citep[e.g.][]{2012ApJ...751L..31B}.

\newpage
\input{ebsample_latex_table}
\clearpage

\begin{deluxetable}{ccrrc}
\tablewidth{0pc}
\tabletypesize{\footnotesize}
\tablecaption{HRS $\&$ APOGEE RV Measurements of KIC 2445134 \label{rv2445134}}
\tablecolumns{5}
\tablehead{
\colhead{UT Date}    & \colhead{BJD-2,400,000} & \colhead{$V_{A}$}   & \colhead{$V_{B}$}      & \colhead{Instrument}  \\
\colhead{}                & \colhead{}                             & \colhead{(km s$^{-1}$)}    & \colhead{(km s$^{-1}$)} & \colhead{} 
}   
\startdata
2011 Aug 27 	 & 	55800.804144 	 & 	 	 9.614  $\pm$ 0.0450 	 & 	 \nodata              	  &  HRS	       \\
2011 Sep 06 	 & 	55810.756457 	 & 	 	 48.652 $\pm$ 0.056 	 & 	 $-$45.01 $\pm$ 3.24        &  HRS	       \\    
2011 Sep 11 	 & 	55815.724269 	 & 	 	$-$15.022 $\pm$ 0.053 	 & 	 \nodata              	  &  HRS	        \\ 
2011 Sep 24 	 & 	55828.729442 	 & 	 	 57.916 $\pm$ 0.058 	 & 	 $-$68.20  $\pm$  3.66    &  HRS	        \\
2011 Oct 21 	 & 	55855.631824 	 & 	 	 28.320 $\pm$ 0.050 	 & 	 \nodata              	  &  HRS	       \\
2011 Oct 29 	 & 	55863.602364 	 & 	 	 39.702 $\pm$ 0.056 	 & 	 \nodata              	  &  HRS	          \\  
2011 Sep 07 	 & 	55811.613072 	 & 	 	 58.55  $\pm$ 0.290	  	 &   $-$67.79  $\pm$  1.37    &  APOGEE	        \\ 
2011 Oct 06 	 & 	55840.593338 	 & 	 	$-$12.33  $\pm$ 0.22	 &    101.92   $\pm$  0.89    &  APOGEE	       \\ 
2011 Oct 17 	 & 	55851.578523 	 & 	 	 17.48  $\pm$ 0.200	  	 &    28.34    $\pm$  0.81    &  APOGEE	       \\ 
\enddata
\end{deluxetable}

\begin{deluxetable}{ccrrc}
\tablewidth{0pc}
\tabletypesize{\footnotesize}
\tablecaption{HRS $\&$ APOGEE RV Measurements of KIC 3003991 \label{rv3003991}}
\tablecolumns{5}
\tablehead{
\colhead{UT Date}    & \colhead{BJD$-$2,400,000} & \colhead{$V_{A}$}   & \colhead{$V_{B}$}      & \colhead{Instrument}  \\
\colhead{}                & \colhead{}                             & \colhead{(km s$^{-1}$)}    & \colhead{(km s$^{-1}$)} & \colhead{} 
}   
\startdata
 2011 Oct 08	 &  55842.687885	 &  $-$144.119 $\pm$ 0.072	 &  $-$50.68 $\pm$ 8.35     &  HRS	        \\
 2012 Mar 21	 &  56007.990683	 & 	$-$120.157 $\pm$ 0.065	 &  \nodata              &  HRS	            \\    
 2012 Apr 24	 &  56041.897439	 & 	$-$101.114 $\pm$ 0.064	 &  $-$201.51 $\pm$ 7.01   &  HRS	         \\ 
 2012 May 28	 &  56075.798304	 & 	$-$143.297 $\pm$ 0.061	 &  $-$50.92  $\pm$ 8.21   &  HRS	         \\
 2012 Jun 08	 &  56086.797501	 & 	$-$103.329 $\pm$t 0.064	 &  $-$185.18 $\pm$ 12.08  &  HRS	        \\
 2012 Jun 10	 &  56088.766870	 & 	$-$140.972 $\pm$ 0.06	     &  $-$54.93  $\pm$ 5.63   &  HRS	       \\  
 2011 Sep 09	 &  55813.6973400	 & 	$-$144.110  $\pm$ 0.290	     &  $-$39.99  $\pm$ 4.58    &  APOGEE        \\ 
 2011 Sep 19	 &  55823.7244600	 & 	$-$115.150  $\pm$ 0.200	     &  $-$151.93 $\pm$ 4.65    &  APOGEE       \\ 
 2011 Oct 06	 &  55840.6597900	 & 	$-$106.210  $\pm$ 0.310	     &  $-$165.85 $\pm$ 6.30    &  APOGEE       \\ 
 2011 Oct 15	 &  55849.5774200	 & 	$-$139.530  $\pm$ 0.200	     &  $-$69.54  $\pm$ 3.22    &  APOGEE        \\ 
 2011 Oct 17	 &  55851.6479100	 & 	$-$136.690  $\pm$ 0.170	     &  $-$72.76  $\pm$ 3.51   &  APOGEE         \\ 
 2011 Nov 01	 &  55866.5693400	 & 	$-$128.360  $\pm$ 0.170	     &  $-$96.77  $\pm$ 4.51    &  APOGEE        \\ 
\enddata
\end{deluxetable}

\begin{deluxetable}{lc}
\tablewidth{0pc}
\tablecaption{KIC 2445134 Orbital \& Physical Parameters\label{orbpar2445134}}
\tablecolumns{2}
\tablehead{\colhead{Parameter} & \colhead{Value}}
\startdata
\multicolumn{2}{c}{\emph{Derived Orbital Parameters}} \\
$P$ (days)  & $8.41201\pm 0.00077$\\
$T_{transit}$ & $2454972.647749$\\
$i$ (deg)   & $88.032\pm0.001$ \\ 
\e         & $0.00555\pm0.00001$ \\
\w{} (rad) & $4.799\pm0.001$ \\
$\Omega_A$ & $15.433\pm0.002$\\
$\Omega_B$ & $18.4\pm0.1$\\ 
$L$3 (\%)  & $4.0\pm0.1$ \\
$T_{peri.}$  & $2455826.85\pm0.25$ \\
\KA{} (\kps) & $37.2\pm0.3$ \\
\KB{} (\kps) & $90\pm1$ \\
\g{} (\kps) & $21.6\pm0.3$ \\
\q{}        & $0.411\pm0.001$ \\ 
$sma$ (\rsun)& $21.3\pm0.2$ \\ \\
$B_A$          & $3.873\pm0.002$ \\ 
$B_B$          & $6.098\pm0.006$ \\ 
\multicolumn{2}{c}{\emph{Physical Parameters}} \\
\MA{} (\msun)  & $1.29\pm0.03$ \\
\MB{} (\msun)  & $0.53\pm0.01$ \\
\RA{} (\rsun)  & $1.42\pm0.01$ \\
\RB{} (\rsun)  & $0.510\pm0.004$ \\
\TB/\TA{}    & $0.635\pm0.001$ \\
$Teff B$ (K) & $3976\pm170$\\ 
\multicolumn{2}{c}{\emph{Gaussian Process Parameters}} \\
$\log$($\phi_\mathrm{matern}$)       & $ -0.91\pm0.05$\\
$\log$($\rho_\mathrm{matern}$)       & $ 5.09\pm0.07$\\
$\log$($\phi_\mathrm{jitter}$)      & $ -9.992\pm0.004$\\
\enddata
\end{deluxetable} 

\begin{deluxetable}{lc}
\tablewidth{0pc}
\tablecaption{KIC 3003991 Orbital \& Physical Parameters\label{orbpar3003991}}
\tablecolumns{2}
\tablehead{\colhead{Parameter} & \colhead{Value}}
\startdata
\multicolumn{2}{c}{\emph{Derived Orbital Parameters}} \\
$P$ (days)  & $7.24478\pm 0.00062$\\
$T_{transit}$ & $2454964.859085$\\
$i$ (deg)   & $88.178\pm0.008$ \\ 
\e         & $0.00030\pm0.00003$ \\
\w{} (rad) & $4.09\pm0.03$ \\
$\Omega_A$ & $18.76\pm0.02$\\
$\Omega_B$ & $19.9\pm0.3$\\ 
$L$3 (\%)  & $3.5\pm0.2$ \\
$T_{peri.}$  & $2455953.65\pm0.18$ \\
\KA{} (\kps) & $24.97\pm0.05$ \\
\KB{} (\kps) & $83\pm1$ \\
\g{} (\kps) & $-122.55\pm0.03$ \\
\q{}        & $0.298\pm0.006$ \\ 
$sma$ (\rsun)& $15.6\pm0.2$ \\ \\
\multicolumn{2}{c}{\emph{Physical Parameters}} \\
\MA{} (\msun)  & $0.74\pm0.04$ \\
\MB{} (\msun)  & $0.222\pm0.007$ \\
\RA{} (\rsun)  & $0.84\pm0.01$ \\
\RB{} (\rsun)  & $0.250\pm0.004$ \\
\TB/\TA{}    & $0.662\pm0.001$ \\
$Teff B$ (K) & $3536\pm140$ \\
\multicolumn{2}{c}{\emph{Gaussian Process Parameters}} \\
$\log$($\phi_{matern}$)       &$-0.3\pm0.1$\\
$\log$($\rho_{matern}$)       &$4.3\pm0.1$\\
$\log$($\phi_{jitter}$)       &$-10.996\pm0.004$\\
\enddata
\end{deluxetable}

\end{document}

%% file: ebsample_latex_table.tex
\begin{longtable}{lllllccl}
\tabletypesize{\scriptsize}\\
\caption{Project EBs Observed \label{ebsampletable}}\\
\hline
\hline
\tablewidth{0pt}
\colhead{KIC ID} & \colhead{$K_p$} & \colhead{$g$} & \colhead{$H$} & \colhead{$P_{\rm{cat}}$} & \colhead{$N_{\rm{APG}}$} & \colhead{$N_{\rm{HET}}$} & \colhead{Notes}\\
\hline
\endfirsthead
\hline
\colhead{KIC ID} & \colhead{$K_p$} & \colhead{$g$} & \colhead{$H$} & \colhead{$P_{\rm{cat}}$} & \colhead{$N_{\rm{APG}}$} & \colhead{$N_{\rm{HET}}$} & \colhead{Notes}\\
\hline
\endhead
\hline
\endfoot
01571511 & 13.424 & 13.855 & 12.041 &  14.022451 &  3 &  9 &                      \\
02010607 & 11.347 & 11.630 & 10.204 &  18.632296 &  3 &  6 &                      \\
02162994 & 14.162 & 14.696 & 12.570 &   4.101595 &  3 &  6 &                      \\
02305372 & 13.821 & 14.341 & 12.201 &   1.404691 &  3 &  0 &                      \\
02305543 & 12.545 & 12.971 & 11.138 &   1.362274 &  3 &  0 &                      \\
02306740 & 13.545 & 14.025 & 12.022 &  10.306987 &  3 &  7 &                      \\
02308957 & 14.520 & 15.031 & 12.916 &   2.219684 &  3 &  6 &                      \\
02309587 & 13.925 & 14.434 & 12.359 &   1.838511 &  3 &  0 &                      \\
02309719 & 12.899 & 13.335 & 11.528 &  54.356360 &  3 &  9 &                      \\
02445134 & 13.551 & 13.948 & 12.201 &   8.412009 &  3 &  8 &                      \\
02445975 & 13.513 & 14.404 & 12.830 &   6.777765 &  3 &  0 &                      \\
02447893 & 14.490 & 15.134 & 12.628 &   0.661620 &  3 &  0 &                      \\
02576692 & 12.744 & 13.191 & 11.371 &  87.878533 &  3 &  8 &                      \\
02583777 & 12.735 & 13.166 & 11.311 &   0.958117 &  3 &  0 &                      \\
02708156 & 10.672 & 10.672 & 10.525 &   1.891272 &  3 &  0 &                      \\
02711114 & 12.335 & 12.634 & 11.115 &   2.858880 &  3 &  0 &                      \\
02720354 & 13.116 & 13.387 & 11.965 &   2.821328 &  3 &  9 &                      \\
02860594 & 13.370 & 13.613 & 12.237 &   5.499945 &  3 & 10 &                      \\
02860788 & 14.043 & 14.614 & 12.255 &   5.259742 &  3 &  0 &                      \\
02997455 & 13.800 & 14.694 & 11.500 &   1.129850 &  3 &  0 & EB Cat FP, No RV Var \\
03003991 & 13.926 & 14.482 & 12.278 &   7.244779 &  6 &  8 &                      \\
03120320 & 10.885 & 11.280 &  9.611 &  10.265613 &  3 &  6 &                      \\
03127817 & 12.155 & 12.242 & 10.590 &   4.327139 &  6 &  9 &                      \\
03128793 & 14.633 & 15.546 & 12.310 &  24.679381 &  6 &  0 &                      \\
03130300 & 14.313 & 14.696 & 12.970 &  11.531282 &  6 &  0 & EB Cat FP, No RV Var \\
03230578 & 13.406 & 13.860 & 12.189 &   6.337611 &  3 &  0 &                      \\
03230787 & 12.553 & 12.990 & 11.141 &  17.734052 &  3 &  8 &                      \\
03241619 & 12.524 & 13.063 & 10.798 &   1.703344 &  3 &  0 &                      \\
03247294 & 13.924 & 14.353 & 12.442 &  67.418828 &  6 &  7 &                      \\
03248033 & 12.161 & 12.427 & 11.019 &   2.668220 &  6 &  0 & EB Cat FP, No RV Var \\
03248332 & 13.102 & 13.369 & 12.012 &   7.363607 &  6 &  6 &                      \\
03335816 & 12.084 & 12.399 & 10.842 &   7.422006 &  3 &  7 &                      \\
03339538 & 13.391 & 14.106 & 11.429 &  14.658014 &  3 &  5 &                      \\
03351945 & 14.734 & 15.549 & 12.605 &   1.080538 &  6 &  0 &                      \\
03352751 & 13.444 & 13.541 & 12.304 &   3.495455 &  6 &  9 &                      \\
03439031 & 11.287 & 11.503 & 10.117 &   5.952026 &  3 &  6 &                      \\
03440230 & 13.636 & 13.706 & 12.486 &   2.881101 &  3 &  0 &                      \\
03441784 &  9.729 &  9.898 &  9.124 &  52.568726 &  3 &  0 &                      \\
03443790 & 11.840 & 12.164 & 10.602 &   1.665784 &  3 &  0 & EB Cat FP, No RV Var \\
03449540 & 14.194 & 14.533 & 12.941 &   3.212006 &  3 &  6 &                      \\
03458919 & 13.815 & 14.370 & 11.512 &   0.892061 &  6 &  0 &                      \\
03541800 & 14.367 & 14.764 & 12.994 &   4.662364 &  3 &  6 & EB Cat FP, No RV Var \\
03542573 & 12.161 & 12.613 & 10.739 &   6.942796 &  3 &  9 &                      \\
03556742 & 14.221 & 14.957 & 12.278 &   0.823013 &  6 &  0 & EB Cat FP, No RV Var \\
03558981 & 13.109 & 13.715 & 11.974 &   2.987858 &  6 &  6 &                      \\
03655326 & 14.213 & 14.614 & 12.980 &  15.066503 &  6 &  0 &                      \\
03656322 & 13.061 & 13.723 & 11.150 &   3.663648 &  6 &  8 &                      \\
03656700 & 12.997 & 13.553 & 11.398 &   0.738528 &  6 &  0 & EB Cat FP, No RV Var \\
03749508 & 13.151 & 13.536 & 11.862 &   1.065734 &  3 &  0 &                      \\
03765771 & 14.216 & 14.601 & 12.897 &   5.567717 &  6 &  0 &                      \\
03766353 & 13.968 & 14.262 & 12.744 &   2.666966 &  6 &  0 &                      \\
03846515 & 12.807 & 13.148 & 11.640 &   1.776084 &  3 &  0 &                      \\
03848919 & 13.901 & 14.477 & 12.141 &   1.047260 &  3 &  0 &                      \\
03848972 & 14.489 & 15.037 & 12.795 &   0.741057 &  3 &  0 &                      \\
03849155 & 13.831 & 14.393 & 12.271 &   1.168313 &  3 &  0 &                      \\
03851193 & 13.682 & 14.050 & 12.478 &   1.341079 &  3 &  7 &                      \\
03858804 & 13.778 & 14.501 & 11.852 &  25.951944 &  6 &  9 & EB Cat FP, No RV Var \\
03858949 & 14.576 & 15.190 & 12.872 &  25.951139 &  6 &  0 & EB Cat FP, No RV Var \\
03861595 & 11.432 & 11.755 & 10.266 &   3.849367 &  6 &  6 &                      \\
03867593 & 13.559 & 13.735 & 12.490 &  73.332022 &  6 &  0 &                      \\
03869825 & 13.320 & 13.597 & 12.180 &   4.800655 &  6 &  8 &                      \\
03955867 & 13.547 & 14.449 & 11.303 &  33.659962 &  3 &  8 &                      \\
03957477 & 12.477 & 13.001 & 10.987 &   0.979052 &  3 &  0 &                      \\
03964562 & 12.403 & 12.419 & 11.903 &   3.012476 &  6 &  0 &                      \\
03965242 & 14.060 & 14.681 & 12.291 &   0.996722 &  6 &  0 &                      \\
03970233 & 14.034 & 14.656 & 12.333 &   8.254914 &  6 &  7 &                      \\
03971315 & 13.664 & 14.039 & 12.320 &   9.892277 &  6 & 11 & EB Cat FP, No RV Var \\
03973549 & 14.293 & 14.801 & 12.926 &   1.389955 &  6 &  0 &                      \\
04069063 & 13.318 & 13.733 & 11.932 &   0.504296 &  6 &  0 &                      \\
04069213 & 12.739 & 13.119 & 11.262 &   5.194256 &  6 &  6 &                      \\
04075064 & 14.951 & 15.712 & 12.904 &  61.422806 &  6 &  0 &                      \\
04076952 & 13.773 & 14.195 & 12.438 &   9.761169 &  3 &  8 &                      \\
04077442 & 13.512 & 14.348 & 11.368 &   0.692843 &  6 &  0 &                      \\
04078693 & 13.485 & 14.131 & 11.794 &   2.756531 &  6 &  0 &                      \\
04157488 & 13.961 & 14.379 & 12.467 &   5.197420 &  3 &  6 &                      \\
04165960 & 13.889 & 14.196 & 12.664 &  13.549178 &  6 &  0 &                      \\
04178389 & 14.228 & 14.695 & 12.708 &  23.210523 &  6 & 10 &                      \\
04275328 & 13.303 & 13.607 & 11.976 &   6.150530 &  6 &  6 & EB Cat FP, No RV Var \\
04281895 & 12.256 & 12.758 & 10.650 &   9.543588 &  6 &  8 &                      \\
04285087 & 12.785 & 13.188 & 11.397 &   4.486031 &  6 &  8 &                      \\
04372379 & 13.810 & 14.091 & 12.689 &   4.535183 &  6 &  7 &                      \\
04376644 & 13.767 & 14.193 & 12.371 &  27.677704 &  6 &  6 &                      \\
04473933 & 12.030 & 12.868 &  9.237 & 103.592625 &  6 &  8 &                      \\
04477830 & 13.548 & 13.894 & 12.336 &   3.384909 &  6 &  6 & EB Cat FP, No RV Var \\
04484356 & 14.235 & 14.872 & 12.359 &   1.144160 &  6 &  0 &                      \\
04570555 & 11.540 & 12.312 &  9.483 &   4.750303 &  6 &  7 & EB Cat FP, No RV Var \\
04570949 & 13.308 & 13.644 & 12.134 &   1.544929 &  6 &  0 &                      \\
04660997 & 12.317 & 12.778 & 10.763 &   0.562561 &  6 &  0 &                      \\
04665989 & 13.016 & 13.215 & 12.000 &   2.248067 &  6 &  0 &                      \\
04671584 & 13.742 & 14.137 & 12.346 &   5.593325 &  6 &  6 &                      \\
04672010 & 14.602 & 15.530 & 12.445 &   0.963042 &  6 &  0 &                      \\
04753561 & 14.928 & 15.918 & 12.647 &   4.944922 &  6 &  0 & EB Cat FP, No RV Var \\
04758368 & 10.805 & 11.670 &  8.516 &   3.749954 &  6 & 12 &                      \\
04840327 & 12.688 & 13.077 & 11.256 &  26.737133 &  6 &  8 &                      \\
04847832 & 12.450 & 13.202 & 11.051 &  30.960237 &  6 &  6 &                      \\
04850874 & 12.228 & 12.379 & 11.037 &   1.775906 &  6 &  0 &                      \\
04851217 & 11.108 & 11.316 & 10.282 &   2.470280 &  6 &  0 &                      \\
04931073 & 11.957 & 12.179 & 10.842 &  26.951236 &  6 &  9 &                      \\
04932691 & 13.627 & 13.815 & 12.641 &  18.112079 &  6 &  0 &                      \\
05017058 & 13.140 & 13.559 & 11.833 &   2.323895 &  6 &  0 &                      \\
05025294 & 13.266 & 13.704 & 11.795 &   5.462690 &  6 &  7 &                      \\
05193386 & 13.998 & 14.703 & 12.054 &  21.378294 &  6 &  8 &                      \\
05199426 & 14.080 & 14.558 & 12.532 &  78.604362 &  6 &  6 &                      \\
05284133 & 12.444 & 12.501 & 11.773 &   8.784576 &  6 &  5 &                      \\
05285607 & 11.419 & 11.684 & 10.304 &   3.899401 &  6 &  7 &                      \\
05288543 & 13.585 & 13.584 & 12.113 &   3.457076 &  6 &  7 &                      \\
05376836 & 14.041 & 14.516 & 12.571 &   3.479425 &  6 &  0 &                      \\
05460835 & 14.293 & 14.720 & 12.855 &  21.539274 &  6 &  0 &                      \\
05462901 & 15.953 & 17.140 & 13.201 &   5.270726 &  6 &  0 &                      \\
\end{longtable}